\documentclass{article}
\usepackage[utf8]{inputenc}
\usepackage{graphicx}
\usepackage[round]{natbib}   
\usepackage{float}
\usepackage{caption}
\usepackage{nicematrix}
\usepackage{subcaption}
\usepackage{todonotes}
\usepackage{rotating}  
\usepackage{amsmath}
\usepackage{algorithm}
\usepackage{algpseudocode}
\usepackage[hidelinks]{hyperref}       
\usepackage{pdflscape}
\usepackage{doi}
\usepackage{graphicx} 
\usepackage{booktabs}
\usepackage{apalike}
\usepackage{amsmath,amsfonts,amssymb,amsthm}
\usepackage{xcolor}
\hypersetup{
    colorlinks,
    linkcolor={red!50!black},
    citecolor={blue!50!black},
    urlcolor={blue!80!black}
}
\usepackage{titling}
\usepackage{lipsum}
\usepackage{multirow}
\usepackage{tabu}
\usepackage{mathtools}
\usepackage[thinlines]{easytable}
\usepackage{array}
\usepackage[page, toc]{appendix} 
\usepackage{afterpage} 

\usepackage{tikz}
\usepackage{bm}
\usepackage{color, colortbl}
\usepackage{hhline}
\usetikzlibrary{arrows,matrix,positioning,fit, decorations.pathreplacing}
\usepackage{titlesec}
\titleformat*{\section}{\large\bfseries}
\titleformat*{\subsection}{\normalsize\bfseries}
\titleformat*{\subsubsection}{\normalsize\bfseries}
\titleformat*{\paragraph}{\normalsize\bfseries}
\titleformat*{\subparagraph}{\normalsize\bfseries}
\usepackage{setspace}
\usepackage{longtable}

\usepackage[left=1in,top=1in,right=1in,bottom=1in,nohead,paperwidth=8.5in, paperheight=11in]{geometry}
\usepackage{authblk}
\theoremstyle{definition}

\ExplSyntaxOn
\prg_new_conditional:Npnn \int_if_zero:n #1 { p , T , F , TF }
  {
    \if_int_compare:w \int_eval:w #1 = \c_zero_int
      \prg_return_true:
    \else:
      \prg_return_false:
    \fi:
  }
\ExplSyntaxOff

\theoremstyle{definition}

\theoremstyle{definition}

\theoremstyle{definition}

\theoremstyle{definition}

\theoremstyle{definition}

\usepackage{url}

\usepackage[normalem]{ulem}
\usepackage{color}

\title{\bf Reduced-Rank Matrix Autoregressive Models: \\ A Medium $N$ Approach}
\author{Alain Hecq, Ivan Ricardo\thanks{Corresponding author: Ivan Ricardo, Maastricht University, School of Business and Economics, Department of Quantitative Economics, P.O.Box 616, 6200 MD Maastricht, The Netherlands. E-mail: iu.ricardo@maastrichtuniversity.nl.}, Ines Wilms}
\affil{Maastricht University, Department of Quantitative Economics}
\date{\today}

\begin{document}
\newtheorem{remark}{Remark}

\maketitle

\begin{abstract}
    Reduced-rank regressions are powerful tools used to identify co-movements within economic time series.
    However, this task becomes challenging when we observe matrix-valued time series, where each dimension may have a different co-movement structure.
    We propose reduced-rank regressions with a tensor structure for the coefficient matrix to provide new insights into co-movements within and between the dimensions of matrix-valued time series.
    Moreover, we relate the co-movement structures to two commonly used reduced-rank models, namely the serial correlation common feature and the index model.
    Two empirical applications involving U.S.\ states and economic indicators for the Eurozone and North American countries illustrate how our new tools identify co-movements.
\end{abstract}



\newpage
\section{Introduction}

In time series econometrics, a quasi-consensus prevails regarding vector autoregressive (VAR) models for analyzing multivariate time series.
Employing VARs in small-scale models, typically characterized by four or five variables, proves suitable for a range of objectives, including (i) testing for Granger causality, (ii) extracting co-movements, (iii) forecasting, and (iv) analyzing the effects of unexpected shocks.
However, when it comes to including many more variables, alternative methodologies must be employed to address the curse of dimensionality.
On the one hand, there exist regularization and sparse regression techniques for large VARs (e.g., \citealp{basu2015regularized, kock2015oracle}).
This strategy has proven to work well in practice and has been used to address points one, three, and four as outlined above (see inter alia \citealp{koop2019hdpanel, nicholson2020forecasting}).
On the other hand, there are dimension reduction techniques such as reduced-rank vector autoregressions (RR-VAR) (see \citealp{cubadda2021survey} for a survey) and factor models (surveyed in \citealp{stock2016dynamic}), which focus more on detecting commonalities, but these may also be used for structural analysis \citep{korobilis2013tvpfactor, carriero2016structural} and forecasting \citep{cubadda2017har, cubadda2019forecasting}. 

For matrix-valued time series, the object of study in this paper, the curse of dimensionality becomes even more prevalent.
Matrix-valued time series entail observing a matrix at each time point instead of a scalar (univariate setting) or a vector (multivariate setting) (see e.g., \citealp{chen2021mar}).
Such matrix-valued data structures are frequently encountered in various contexts.
This includes, amongst others, employment data over different sectors for different U.S.\ states over time \citep{samadi2024marest}, financial characteristics of different companies over time \citep{wang2019factor}, and various air pollutants for different cities over time \citep{yuan2023twowayfactor}.
To make things concrete, consider a four-by-five matrix-valued time series representing four economic indicators for five countries observed over time.
Within the VAR framework, this involves jointly studying 20 variables, leading to a VAR with 400 parameters if one lag is included, or up to 1,200 parameters for three lags; the dimensionality thus quickly explodes.
Although this example is not particularly large, it is a common situation encountered in econometrics which motivates us to explore such ``medium-sized'' matrix-valued time series.
In this paper, we aim to quantify the co-movements not only between the different series but also between different dimensions of the matrix-valued time series.

While it is possible to use dimension reduction or sparse techniques on the vectorized data to study co-movements, doing so overlooks the multidimensional nature of the problem.
In fact, by vectorizing the data before the analysis, the stacked unrestricted VAR cannot distinguish between the two dimensions of the matrix-valued time series and we may therefore lose dimension-specific co-movement interpretations.
For instance, when studying a dynamic financial network \citep{diebold2015network} or an application to international trade \citep{billio2023bayesian}, differences may arise from the global and local effects of importing and exporting countries or the connectedness of different asset returns or volatilities.
By keeping the matrix-valued time series intact, we can thus detect co-movement structures that would not be possible to detect from the vectorized data.

Matrix-valued time series have gained more attention in recent years, notably with the introduction of the matrix autoregressive (MAR) model \citep{chen2021mar}. 
To address the overparameterization in matrix-valued time series models, various extensions have been proposed, including the matrix-valued factor model and its extensions \citep{wang2019factor, chen2020constrained}, tensor factor models \citep{chen2022factor}, and reduced-rank models \citep{xiao2022reduced, wang2024high}.
The literature on factor models encompasses determining the number of factors \citep{han2022determining}, estimation methods \citep{han2022optimal}, and statistical inference \citep{chen2023inference}.
For vector-valued data, methods have emerged that assume a three-dimensional tensor of the VAR coefficient matrix, which entails an $N \times N$ coefficient matrix combined with a third lag dimension, forming an $N \times N \times p$ tensor.
Tensor factorization methods can then be used for simultaneous dimension reduction across the three dimensions separately \citep{wang2022tensorvar, wang2022common}.
Most recently, this idea for a tensor factorization has been extended to matrix-valued time series \citep{wang2024high}, assuming the coefficients form a four-dimensional tensor.

In this paper, we contribute to the literature by bridging the gap between this tensor literature and the dimension reduction literature on reduced-rank structures for VAR models that permit co-movement detection \citep{cubadda2009studying, cubadda2017har, cubadda2021survey, cubadda2022dimension}.
On the one hand, co-movements are typically identified using the serial correlation common feature (SCCF) method \citep{engle1993testing}, where serial correlation is common if a linear combination of the series is no longer serially correlated.
In practice, this entails a reduced-rank structure of the VAR coefficient matrix, or more precisely, an analysis of its left null space.
Since \cite{engle1993testing}, there have been various SCCF extensions, including common trends and cycles \citep{vahid1993common}, the combination of cointegration and SCCF in VAR models \citep{hecq2000permanent}, as well as SCCF applications including non-causal time series \citep{cubadda2019cononcausal}, and envelope models \citep{samadi2023envelope}.
On the other hand, co-movements may be identified by creating ``indexes'', or factors, from a set of data using Index/Factor models \citep{reinsel1983mai}.
This amounts to an analysis of the right null space of the VAR coefficient matrix which allows us to extract a small number of indexes/factors\footnote{It should be noted that both uses of ``index'' and ``factor'' are possible in this case. However, to keep consistency with the current literature, we continue with the latter denomination.} from linear combinations of the original variables.
This method of extracting factors from the data is closely related to the dynamic factor model of \cite{lam2012factor}, where the errors are assumed to be serially uncorrelated and the factors follow a dynamic structure.
Moreover, this methodology has enjoyed success in topics including the analysis of structural shocks \citep{carriero2016structural}, forecasting \citep{cubadda2017har, cubadda2019forecasting}, and cointegration \citep{cubadda2024vecm}.
Recently, \cite{cubadda2022dimension} link the framework of both the SCCF method and the factor model for VAR models by assuming a common left and right null space of the VAR coefficient matrix.
Their work inspired us to provide a reduced-rank matrix autoregressive model that can link the two methodologies, SCCF and factor models, without assuming a common left and right null space.

Our paper offers the following contributions.
First, our matrix-valued reduced-rank time series model is notably more parsimonious and informative about economic nexus than a stacked VAR model under reduced-rank restrictions.
Matrix-valued time series are infamous for having many parameters to estimate, thus this efficiency is crucial in capturing the underlying dynamics of the data while avoiding unnecessary complexity.
Second, by employing a tensor structure on the coefficients in the matrix autoregressions, we can discern different sources of co-movements.
To this end, we extend the work by \cite{wang2024high} to multiple lags thereby allowing the factor structure to vary along each lag.
Third, our modeling framework enables us to identify whether co-movements arise from a factor structure, a common feature structure characterized by serial correlation, or a combination of both.
This is achieved without assuming the same rank among the row and column spaces of the VAR coefficient matrix as done in \cite{cubadda2022dimension}, but rather, by allowing these two spaces to differ.
Finally, our study permits exploring full-rank models in certain dimensions, a strategy not accommodated by the large $N$ approach of \cite{lam2012factor} or \cite{wang2024high}, yet key to allow for in our medium-sized setup.
Consequently, we employ information criteria within a four-dimensional approach to detect potential left and right null space reductions at both dimensions of the matrix-valued time series.

The remainder of this paper is structured as follows. Section \ref{sec:representation} starts by reviewing preliminaries on tensors before discussing their usage in the context of our proposed reduced-rank matrix autoregressive (RR-MAR) model. 
Section \ref{sec:estimationselection} details the estimation of the RR-MAR model, and how we select the ranks of the coefficient tensor.
Section \ref{sec:simulation} contains a simulation study where we evaluate the performance of the proposed rank selection method.
Section \ref{sec:empirical} gives two examples of our method applied to coincident and leading indicators of U.S.\ states and various economic indicators of different countries.
Section \ref{sec:conclusion} concludes and discusses future research avenues.

A word on notation.
Throughout this paper, we denote vectors by boldface small letters $\mathbf{x}$, matrices by boldface capital letters $\mathbf{X}$, and tensors by calligraphic letters $\mathcal{X}$. 
For a generic matrix $\mathbf{X}$, we call  $\mathbf{X}^\top$, $\|\mathbf{X}\|_F$, $\text{vec}(\mathbf{X})$, respectively the transpose, Frobenius norm, and column-wise vectorization.

\section{Model Representation}
\label{sec:representation}
We begin by introducing in Section \ref{sec:notations} preliminary notations and the Tucker decomposition of a tensor, which is key to understanding the reduced-rank structure of the coefficient tensor.
In Section \ref{sec:rrmarmodel}, we then introduce the Reduced-Rank Matrix Autoregressive (RR-MAR) model, along with its relation to various reduced-rank VAR models in the literature, namely the SCCF representation and Index/Factor model, in Section \ref{sec:reducedrankrelation}.

\subsection{Preliminaries and Tucker Decomposition}
\label{sec:notations}

We start by reviewing the basic notation used in tensor analysis. 
For a comprehensive treatment of tensor algebra and decompositions, readers can refer to \citet{kolda2009tensor}.
Intuitively, a tensor is a generalization of vectors and matrices. 
Just as we can create a matrix from the outer product of two vectors, we can construct a three-dimensional array or tensor of data from the product of three vectors.
This concept can be extended to multi-dimensional objects, which are $K$-dimensional arrays or tensors.
We typically refer to $K$ as the number of dimensions, but this can interchangeably be called the order, or mode of the tensor.
Scalars correspond to 0-order tensors, vectors to 1-order tensors, and matrices to 2-order tensors.
For higher-order tensors (tensors with more than 2 dimensions), we label the indices as $N_1, N_2, \dots, N_K$ where $K$ is the number of modes of a tensor.
We label the elements of a $K$-order real-valued tensor $\mathcal{Y} = (\mathcal{Y}_{n_1, \dots, n_K}) \in \mathbb{R}^{N_1 \times \dots \times N_K}$ with entries $\mathcal{Y}_{n_1, \dots, n_K}$ for $n_k = 1, \dots N_k$ and $k = 1, \dots, K$.
To ease understanding, we illustrate the tensor notation throughout this section for an example with four economic indicators and five countries as in our empirical application in Section \ref{sec:empirical}.
By including time as an additional dimension, we can thus create a third-order tensor, $\mathcal{Y} \in \mathbb{R}^{N_1 \times N_2 \times T}$, where $N_1 = 4$ (indicators), $N_2 = 5$ (countries), and $N_3 = T$ (time).

Matricization, also known as unfolding or flattening, is the process of reordering the elements of a third- or higher-order tensor into a matrix. 
For any $K$-dimensional tensor $\mathcal{Y} \in \mathbb{R}^{N_1 \times \dots \times N_K}$, we denote its mode-$k$ unfolding $\mathbf{Y}_{(k)} \in \mathbb{R}^{N_k \times N_{-k}}$, with $N_{-k}:= \prod_{i = 1, i \neq k}^K N_i$, as the matrix obtained by setting the $k$-th tensor mode as its rows and collapsing all the other dimensions into the columns.
Specifically, the $(n_1, \dots, n_K)$-th element of $\mathcal{Y}$ is mapped to the $(n_k, j)$-element of $\mathbf{Y}_{(k)}$, where 
\[
j = 1 + \sum_{\substack{l = 1,\\l \neq k}}^K (n_l - 1) J_l \quad \text{with} \quad J_l = \prod_{\substack{m = 1,\\m \neq k}}^{l - 1} N_m.
\]
To matricize our four-by-five-by-time example into vector-valued data, appropriate for usage in a VAR setup, we would unfold it along the third dimension.
This would yield a $T \times 20$ matrix $\mathbf{Y}_{(3)}$.
If we also wish to determine to which element $\mathcal{Y}_{3,2, T}$ (i.e.\ value of the third indicator for country two at time $T$) is mapped, the formula above 
would yield $j = 1 + (n_1 - 1) + (n_2 - 1) (N_1) = 7$ where $n_1 = 3$, $n_2 = 2$, and $N_1 = 4$.
Thus, this element would map to the $(T, 7)$-th element of $\mathbf{Y}_{(3)}$.

This one-mode matricization can be generalized to include multiple modes associated with the rows of the matricized tensor.
For this, we follow \cite{wang2024high} in their definition of the generalized multi-mode matricization.
For any index subset $S \subset \{1, \dots, K\}$, the multi-mode matricization $\mathbf{X}_{[S]}$ is the $\prod_{i \in S} N_i$-by-$\prod_{i \notin S} N_i$ matrix whose $(i,j)$-th element is mapped from the $(n_1, \dots, n_K)$-th element of $\mathcal{X}$, where
\[
i = 1 + \sum_{l \in S}(n_l - 1) I_l \quad \text{and} \quad j = 1 + \sum_{l \notin S}(n_l - 1) J_l \quad \text{with} \quad I_l = \prod_{\substack{m \in S,\\m < k}} N_m \quad \text{and} \quad J_l = \prod_{\substack{m \in S,\\m < k}} N_m.
\]

With the concept of matricization established, we can extend matrix multiplication to tensors as follows.
Given any tensor $\mathcal{Y} \in \mathbb{R}^{N_1 \times \dots \times N_K}$ and matrix $\mathbf{U} \in \mathbb{R}^{N_k \times M}$, we denote the $k$-mode product as $\mathcal{Y} \times_k \mathbf{U}$ and produces a $K$-th order tensor in $\mathbb{R}^{N_1 \times \dots \times N_{k-1} \times M \times N_{k+1} \times \dots \times N_K}$ defined element-wise as 
\[
(\mathcal{Y} \times_k \mathbf{U})_{n_1 \dots n_{k-1} m n_{k+1} \dots n_K} = \mathbf{U} \mathbf{Y}_{(k)} = \sum_{n_k = 1}^{N_k} \mathcal{Y}_{n_1 \dots n_K} \mathbf{U}_{m n_k}.
\]

Furthermore, consider the multi-mode product, similar to the sequence of contracted product outlined in \cite{billio2023bayesian} to multiply over multiple tensor modes.
For any two tensors $\mathcal{X} \in \mathbb{R}^{M_1 \times \dots \times M_L \times N_1 \times \dots \times N_K}$ and $\mathcal{Y} \in \mathbb{R}^{N_1 \times \dots \times N_K \times P_1 \times \dots \times P_Q}$, we denote this operation as $\mathcal{X} \bar{\times}_K \mathcal{Y}$.
The result is an $(L+Q)$-order tensor in $\mathbb{R}^{M_1 \times \dots \times M_L \times P_1 \times \dots \times P_Q}$ defined element-wise as
\[
(\mathcal{X} \bar{\times}_K \mathcal{Y})_{m_1 \dots m_L p_1 \dots p_Q} = \sum_{n_1 = 1}^{N_1} \dots \sum_{n_K = 1}^{N_K} \mathcal{X}_{m_1 \dots m_L n_1 \dots n_K} \mathcal{Y}_{n_1 \dots n_K p_1 \dots p_Q}.
\]
Returning to our four-by-five-by-time tensor example, we may wish to multiply this with a four-dimensional coefficient tensor $\mathcal{A} \in \mathbb{R}^{4 \times 5 \times 4 \times 5}$ over the last two dimensions, as typically required in our RR-MAR model framework.
This is feasible because the last two dimensions of $\mathcal{A}$ precisely match the first two dimensions of $\mathcal{Y}$.
Consequently, the result of $\mathcal{A} \bar{\times}_2 \mathcal{Y}$ yields a new tensor $\mathcal{Y}^* \in \mathbb{R}^{4 \times 5 \times T}$.

\begin{figure}
    \centering

\tikzset{every picture/.style={line width=0.75pt}} 
\begin{tikzpicture}[x=0.75pt,y=0.75pt,yscale=-1,xscale=1]

\draw  [fill={rgb, 255:red, 100; green, 143; blue, 255 }  ,fill opacity=1 ] (203.45,165.39) -- (294.27,165.39) -- (294.27,248.78) -- (203.45,248.78) -- cycle ;

\draw  [fill={rgb, 255:red, 100; green, 143; blue, 255 }  ,fill opacity=1 ] (54.1,94.39) -- (136.58,94.39) -- (136.58,177.77) -- (54.1,177.77) -- cycle ;
\draw  [fill={rgb, 255:red, 220; green, 38; blue, 127 }  ,fill opacity=1 ] (169.67,61.75) -- (136.58,94.39) -- (54.1,94.39) -- (87.52,62.37) -- cycle ;
\draw  [fill={rgb, 255:red, 255; green, 176; blue, 0 }  ,fill opacity=1 ] (168.85,62.57) -- (169.22,144.68) -- (135.75,178.6) -- (135.75,95.21) -- cycle ;
\draw  [fill={rgb, 255:red, 255; green, 176; blue, 0 }  ,fill opacity=1 ] (361.63,70.04) -- (362,152.15) -- (328.53,186.07) -- (328.53,102.68) -- cycle ;

\draw  [fill={rgb, 255:red, 220; green, 38; blue, 127 }  ,fill opacity=1 ] (331.49,20.47) -- (298.4,53.11) -- (215.92,53.11) -- (249.34,21.09) -- cycle ;

\draw  [fill={rgb, 255:red, 120; green, 94; blue, 240 }  ,fill opacity=1 ] (227.67,100.37) -- (245.07,82.97) -- (285.67,82.97) -- (285.67,124.18) -- (268.27,141.58) -- (227.67,141.58) -- cycle ; \draw   (285.67,82.97) -- (268.27,100.37) -- (227.67,100.37) ; \draw   (268.27,100.37) -- (268.27,141.58) ;

\draw (179.63,98.82) node [anchor=north west][inner sep=0.75pt]  [font=\LARGE]  {$\approx $};
\draw (240.04,110.38) node [anchor=north west][inner sep=0.75pt]  [font=\LARGE]  {$\mathcal{G}$};
\draw (330,120.94) node [anchor=north west][inner sep=0.75pt]  [font=\LARGE]  {$\mathbf{U}_{2}$};
\draw (257.33,26.82) node [anchor=north west][inner sep=0.75pt]  [font=\LARGE]  {$\mathbf{U}_{3}$};
\draw (80,125) node [anchor=north west][inner sep=0.75pt]  [font=\LARGE]  {$\mathcal{A}$};
\draw (235.04,196.9) node [anchor=north west][inner sep=0.75pt]  [font=\LARGE]  {$\mathbf{U}_{1}$};

\end{tikzpicture}
    \caption{Tucker decomposition of a three-dimensional tensor $\mathcal{A}$. $\mathcal{G}$ is a compressed core tensor, while $\mathbf{U}_i$ for $i = 1, 2, 3$ are factor matrices for each dimension.}
    \label{fig:tucker}
\end{figure}
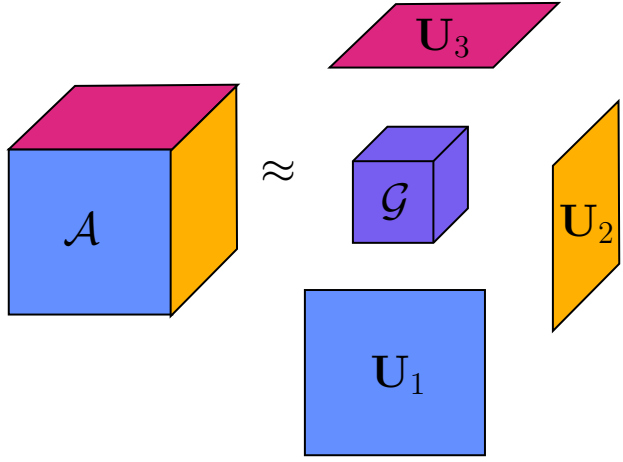

Next, to define low rankness in a tensor, we employ the Tucker decomposition \citep{tucker1966decomposition, delathauwer2000best}, which considers the rank of the matricized tensor across all modes.
For any tensor $\mathcal{A} \in \mathbb{R}^{N_1 \times \dots \times N_K}$, the Tucker ranks $(r_1, \dots, r_K)$ are defined as the matrix ranks of each one-mode matricization, denoted as $r_i = \text{rank}(\mathbf{A}_{(i)})$, for $i = 1, \dots, K$.
Note that these ranks are analogous to column and row ranks for a matrix, but may vary for higher-order tensors.
Suppose the tensor $\mathcal{A} \in \mathbb{R}^{N_1 \times N_2 \times \dots \times N_K}$ has Tucker ranks $(r_1, r_2, \dots, r_K)$.
Then a Tucker decomposition for $\mathcal{A}$ can be expressed as follows
\begin{equation}
    \label{eq:tuckerdecomp}
    \mathcal{A} \approx [[\mathcal{G}; \mathbf{U}_1, \mathbf{U}_2, \dots, \mathbf{U}_K]] = \mathcal{G} \times_1 \mathbf{U}_{1} \times_2 \mathbf{U}_{2} \times_3 \dots \times _K \mathbf{U}_{K},
\end{equation}
where $\mathcal{G} \in \mathbb{R}^{r_1 \times r_2 \times \dots \times r_K}$ is the core tensor, and $\mathbf{U}_{i} \in \mathbb{R}^{N_i \times r_i}$ for $i = 1, \dots, K$ are the factor matrices.
Figure \ref{fig:tucker} visualizes this decomposition in three dimensions.
The Tucker decomposition can be equivalently represented in a flattened form.
Specifically, by considering any $k$-mode matricization, we can express the Tucker decomposition as
\[
\mathbf{A}_{(k)} = \mathbf{U}_{k} \mathbf{G}_{(k)} \left(\mathbf{U}_{K} \otimes \dots \otimes \mathbf{U}_{k+1} \otimes \mathbf{U}_{k-1} \otimes \dots \otimes \mathbf{U}_{1}\right)^\top,
\]
where $\otimes$ denotes the Kronecker product.
Similarly, a multi-mode matricization yields two groupings of factor matrices along with the core tensor, as shown by
\[
\mathbf{A}_{[S]} = \left(\otimes_{i \in S} \mathbf{U}_{i}\right) \mathbf{G}_{[S]} \left(\otimes_{i \notin S} \mathbf{U}_{i}\right)^\top,
\]
where $\otimes_{i \in S}$ and $\otimes_{i \notin S}$ are Kronecker products operating in the reverse order within the corresponding index sets.

One caveat of the Tucker decomposition is a condition on the ranks, namely, the Tucker ranks must satisfy the condition
\begin{align} \nonumber
    r_i \leq \min (N_i, \prod_{\substack{j = 1,\\j \neq i}}^{K} r_j).
\end{align}
In other words, the maximum Tucker rank of a tensor must not exceed the product of the other ranks.
This condition stems from a rank restriction on matrices, where the rank of a matrix must be less than or equal to the matrix's smallest dimension.

Finally, note that the Tucker decomposition in equation \eqref{eq:tuckerdecomp} is not unique since 
\[
[[\mathcal{G}; \mathbf{U}_1, \mathbf{U}_2, \dots, \mathbf{U}_K]] = [[\mathcal{G} \times_1 \mathbf{O}_1 \times_2 \mathbf{O}_2 \times_3 \dots \times_K \mathbf{O}_K; \mathbf{U}_1 \mathbf{O}_1^{-1}, \mathbf{U}_2 \mathbf{O}_2^{-1}, \dots, \mathbf{U}_K \mathbf{O}_K^{-1}]],
\]
for any nonsingular orthogonal matrices $\mathbf{O}_1 \in \mathbb{R}^{N_1 \times r_1}, \mathbf{O}_2 \in \mathbb{R}^{N_2 \times r_2}, \dots, \mathbf{O}_K \in \mathbb{R}^{N_K \times r_K}$.
To circumvent this, we follow the standard practice of considering a special Tucker decomposition: the higher-order singular value decomposition (HOSVD) \citep{delathauwer2000hosvd}.
The HOSVD considers all factor matrices $\mathbf{U}_{i}$ to be orthogonal consisting of the top $r_i$ left singular vectors of $\mathbf{A}_{(i)}$, and the core tensor has the property of so-called ``all-orthogonality'', namely, $\mathbf{G}_{(i)} \mathbf{G}_{(i)}^\top$ is diagonal.
This restriction is sufficient to ensure that the Tucker decomposition is rotationally invariant, however, additional restrictions may be needed to ensure scale and sign indeterminacies. 

\subsection{Reduced-Rank Matrix Autoregressive Model}
\label{sec:rrmarmodel}

We present a low-rank matrix autoregressive model for matrix-valued time series and call this the Reduced-Rank Matrix Autoregressive (RR-MAR) model.
This framework extends the approach of \citet{wang2024high} toward the identification of serial correlation common features and factor models.
For simplicity, we begin with the model containing a single lag but expand this special case to multiple lags in the latter half of this section.
Our model in tensor form is given by
\begin{align}
    \label{eq:mar}
    \mathbf{Y}_t = \mathcal{A} \bar{\times}_2 \mathbf{Y}_{t-1} + \mathbf{E}_t,
\end{align}
where $\mathbf{Y}_t \in \mathbb{R}^{N_1 \times N_2}$ is the response matrix, $\mathcal{A} \in \mathbb{R}^{N_1 \times N_2 \times N_1 \times N_2}$ is a four-dimensional coefficient tensor, and $t = 1, \dots, T$.
Additionally, $\mathbf{E}_t$ is an $N_1 \times N_2$ matrix-valued error term such that $E(\text{vec}(\mathbf{E}_t)) = \bm{0}$, $E(\text{vec}(\mathbf{E}_t) \text{vec}(\mathbf{E}_t)^\top) = \Sigma_{\mathbf{E}}$ is a finite and positive definite matrix, $E(\text{vec}(\mathbf{E}_t) | \mathcal{F}_{t-1}) = \bm{0}$ and $\mathcal{F}_t$ is the natural filtration of the process $\mathbf{Y}_t$.
For stationarity of the model, we require that the determinant of $\mathbf{A}_{[2]}(z)$ is not equal to zero for all $z \in \mathbb{C}$ and $|z| < 1$, where $\mathbf{A}_{[2]}(z) = \mathbf{I}_{N_1 N_2} - \sum_{j=1}^p \mathbf{A}_{[2], j} z^j$ is the VAR matrix polynomial and $\mathbf{I}_{N_1 N_2}$ is the $N_1 N_2 \times N_1 N_2$ identity matrix.
To simplify our analysis, we assume that the data have been properly demeaned to account for the constant term.
Finally, we assume the coefficient tensor has a Tucker decomposition given by
\begin{align*}
    \mathcal{A} = \mathcal{G} \times_1 \mathbf{U}_{1} \times_2 \mathbf{U}_{2} \times_3 \mathbf{U}_{3} \times_4 \mathbf{U}_{4},
\end{align*}
with $\mathcal{G} \in \mathbb{R}^{r_1 \times r_2 \times r_3 \times r_4}, \mathbf{U}_{1} \in \mathbb{R}^{N_1 \times r_1}$, $\mathbf{U}_{2} \in \mathbb{R}^{N_2 \times r_2}$, $\mathbf{U}_{3} \in \mathbb{R}^{N_1 \times r_3}$, $\mathbf{U}_{4} \in \mathbb{R}^{N_2 \times r_4}$.
The factor matrices, $\mathbf{U}_i$ for $i = 1, \dots 4$ control the rank of each dimension, while the core tensor $\mathcal{G}$ allows us to combine potentially different rank combinations into our coefficient tensor.
In our example with four economic indicators and five countries, the $\mathcal{G}$ core tensor would depend on the selected ranks and the factor matrices would be of dimension $\mathbf{U}_1 \in \mathbb{R}^{4 \times r_1}$, $\mathbf{U}_2 \in \mathbb{R}^{5 \times r_2}$, $\mathbf{U}_3 \in \mathbb{R}^{4 \times r_3}$, $\mathbf{U}_4 \in \mathbb{R}^{5 \times r_4}$ for fixed Tucker ranks $r_1, \dots, r_4$.

The RR-MAR model in equation \eqref{eq:mar} can be related to the VAR model.
To this end, we vectorize the matrix-valued time series to obtain
\begin{align}
    \label{eq:rrmar1}
    \text{vec}(\mathbf{Y}_t) &= (\mathbf{U}_{2} \otimes \mathbf{U}_{1}) \mathbf{G}_{[2]} (\mathbf{U}_{4} \otimes \mathbf{U}_{3})^\top \text{vec}(\mathbf{Y}_{t-1}) + \text{vec}(\mathbf{E}_t),
\end{align}
with $\mathbf{A}_{[2]} = (\mathbf{U}_{2} \otimes \mathbf{U}_{1}) \mathbf{G}_{[2]} (\mathbf{U}_{4} \otimes \mathbf{U}_{3})^\top \in \mathbb{R}^{N_1 N_2 \times N_1 N_2}$ being the VAR($1$) coefficient matrix.
Equation \eqref{eq:rrmar1} allows us to see the exact restrictions we place on the VAR model, namely a decomposition of the coefficient matrix into three matrices, two of which have a Kronecker structure.
The corresponding VAR coefficient rank can be determined from the Tucker ranks as $\min (r_1 r_2, r_3 r_4)$.

\begin{remark}
    Model \eqref{eq:rrmar1} can be linked to the Reduced-Rank Autoregression for matrix-valued time series of \cite{xiao2022reduced}.
    Here, the authors propose a similar Kronecker structure on the coefficients of the VAR model but omit the presence of the core tensor $\mathcal{G}$.
    In practice, this boils down to a special case of our model where $r_1 = r_3$ and $r_2 = r_4$.
    Under this assumption and the exclusion of the core tensor, the vectorized model \eqref{eq:rrmar1} becomes
    \begin{align*}
        \text{vec} (\mathbf{Y}_t) &= (\mathbf{U}_2 \otimes \mathbf{U}_1) (\mathbf{U}_4 \otimes \mathbf{U}_3)^\top \text{vec} (\mathbf{Y}_{t-1}) + \text{vec}(\mathbf{E}_t), \\
        \text{vec} (\mathbf{Y}_t) &= (\mathbf{U}_2 \mathbf{U}_4^\top \otimes \mathbf{U}_1 \mathbf{U}_3^\top) \text{vec} (\mathbf{Y}_{t-1}) + \text{vec}(\mathbf{E}_t), \\
        \mathbf{Y}_t &= \mathbf{U}_1 \mathbf{U}_3^\top \mathbf{Y}_{t-1} \mathbf{U}_4 \mathbf{U}_2^\top + \mathbf{E}_t,
    \end{align*}
    which corresponds to the model from \cite{xiao2022reduced}.
    
    We thus generalize the latter by including a core tensor, thereby permitting interpretations in terms of SCCF and factor models while maintaining the flexibility to have different ranks associated with each factor matrix.
\end{remark}

\begin{remark}
    Although we encompass various reduced-rank models, our method is more restrictive than the traditional RR-VAR.
    To see this, note that the RR-VAR(1) has the form
    \begin{align*}
        \mathbf{y}_t = \mathbf{A}\mathbf{B}^\top \mathbf{y}_{t-1} + \mathbf{e}_t,
    \end{align*}
    whereas the RR-MAR($1$) is given by equation \eqref{eq:rrmar1}.
    The RR-MAR($1$) thus decomposes both the $\mathbf{A}$ and $\mathbf{B}$ into a Kronecker structure as in \cite{xiao2022reduced}.
    This Kronecker structure allows for groupings of the different dimensions, with the factor matrices $\mathbf{U}_1$ and $\mathbf{U}_2$ holding dimension-specific interpretations for the $\mathbf{A}$ matrix, and $\mathbf{U}_3$ and $\mathbf{U}_4$ holding dimension-specific interpretations for the $\mathbf{B}$ matrix.
    This restriction is explored in depth in Section \ref{sec:reducedrankrelation}. 
\end{remark}

To extend the RR-MAR model in equation \eqref{eq:rrmar1} with one lag to $p$ lags, we impose that the factor matrices remain the same across all lags while allowing the core tensor to vary with each lag.
This approach preserves the interpretability of the SCCF and factor models across the different lags and provides the flexibility to accommodate different ranks between the two interpretations.
The RR-MAR in tensor and vector form can then be expressed as
\begin{align}
    \label{eq:rrmarp}
    \mathbf{Y}_t &= \sum_{i=1}^p \mathcal{A}_i \bar{\times}_2 \mathbf{Y}_{t-i} + \mathbf{E}_t, \nonumber \\
     \Leftrightarrow \text{vec}(\mathbf{Y}_t) &= \sum_{i=1}^p (\mathbf{U}_{2} \otimes \mathbf{U}_{1}) \mathbf{G}_{[2], i} (\mathbf{U}_{4} \otimes \mathbf{U}_{3})^\top \text{vec}(\mathbf{Y}_{t-i}) + \text{vec}(\mathbf{E}_t),
\end{align}
where $\mathcal{A}_i = \mathcal{G}_i \times_1 \mathbf{U}_{1} \times_2 \mathbf{U}_{2} \times_3 \mathbf{U}_{3} \times_4 \mathbf{U}_{4} \in \mathbb{R}^{N_1 \times N_2 \times N_1 \times N_2}$ is a four-dimensional coefficient tensor across the different lags of the matrix-valued time series, and $\mathbf{G}_{[2], i}$ is the corresponding matricized core tensor over the first two dimensions.
An alternative representation is a ``stacked'' version of equation \eqref{eq:rrmarp}, where the lagged terms $\mathbf{Y}_{t-i}$ are concatenated into a new tensor $\mathcal{X}_t \in \mathbb{R}^{N_1 \times N_2 \times p}$.
The stacked vectorized RR-MAR($p$) would then be
\begin{align}
    \label{eq:stackedrrmarp}
    \text{vec}(\mathbf{Y}_t) = (\mathbf{U}_{2} \otimes \mathbf{U}_{1}) \mathbf{G}_{[2]}^* (\mathbf{I}_p \otimes \mathbf{U}_{4} \otimes \mathbf{U}_{3})^\top \text{vec}(\mathcal{X}_{t}) + \text{vec}(\mathbf{E}_t),
\end{align}
where $\mathbf{G}_{[2]}^* \in \mathbb{R}^{r_1 r_2 \times r_3 r_4 p}$ is a wide core tensor holding the lagged slope terms and $\mathbf{I}_p$ is the $p \times p$ identity matrix.
The inclusion of the identity matrix is commonly done in the index model literature \citep{velu1986reduced, carriero2016structural}, and can be used to extract factors by pre- and post-multiplying each lag to the matrix-valued series $\mathbf{U}_3^\top \mathbf{Y}_{t-p} \mathbf{U}_4 = \mathbf{F}_{t-p} \in \mathbb{R}^{r_3 \times r_4}$ for $p = 1, \dots, P$.
Thus, this representation of the RR-MAR allows for both factor and SCCF interpretations, where the $\mathbf{U}_1$ and $\mathbf{U}_2$ factor matrices can extract SCCF vectors, while $\mathbf{U}_3$ and $\mathbf{U}_4$ can extract factors (see Sections \ref{sec:sccf} and \ref{sec:factormodels}).

Finally, compared to the RR-VAR or the traditional factor model, note that the RR-MAR($p$) model requires the estimation of considerably fewer parameters.
The total number of parameters in the RR-MAR is given by
\begin{equation}
    \label{eq:tuckpars}
    \prod_{i=1}^{4} r_i p + \sum_{i=1}^2 r_i (N_i - r_i) + r_{2+i} (N_i - r_{2+i}),
\end{equation}
where we account for the degrees of freedom imposed by the orthogonality constraints as further discussed in Section \ref{sec:notations}.
Here, the first product gives us the number of parameters in the core tensor, while the subsequent sum denotes the number of parameters in each $\mathbf{U}_{i}$ factor matrix.
Moreover, in the scenario where $r_1 = r_3 = N_1$ or $r_2 = r_4 = N_2$ (full rank case), the factor matrices are no longer present, leaving the core tensor to contain all the VAR parameters.

\begin{remark}
    The RR-MAR($p$) model in equation \eqref{eq:rrmarp} can be related to the multilinear low-rank vector autoregressive models of \cite{wang2022tensorvar} and tensor model of \cite{wang2024high}.
    While \cite{wang2024high} explores the possibility of including lags in a tensor autoregressive model, they neither formalize this nor discuss its practical implications. 
    In contrast, our model accommodates multiple lags and offers direct insights into the factors extracted from it.
    \cite{wang2022tensorvar} discusses reducing the lag dimension, however, they focus solely on vector-valued time series, assuming the VAR coefficient to be a tensor of size $N \times N \times p$.
    Our model bridges the gap between these methods for vector-valued and tensor-valued data by incorporating lags in matrix-valued time series, emphasizing our interest in reduced-rank structures that persist among all lags.
    
\end{remark}

\subsection{Relation to Reduced-Rank VAR Models}
\label{sec:reducedrankrelation}

We now relate the RR-MAR to popular reduced-rank models, namely, the serial correlation common feature representation and the factor model.

\subsubsection{Serial Correlation Common Features}
\label{sec:sccf}

In the RR-VAR framework, the serial correlation common feature (SCCF) model \citep{engle1993testing} involves using the left null space from the coefficient matrix of the VAR to discern the contemporaneous co-movements between different series.
This null space provides information on the co-movement relations between the series in the sample, and to what extent one series scales with respect to another.
For matrix-valued time series, we encounter two dimensions related to the SCCF method: one for the row space and another for the column space of the matrix observed over time. 
The major distinction between the proposed RR-MAR and the traditional RR-VAR is the potential for each of the two dimensions to have a distinct rank.

Starting from our RR-MAR(1) in equation \eqref{eq:rrmar1} we identify orthogonal complements, $\bm{\delta} \in \mathbb{R}^{N_1 \times (N_1 - r_1)}$ and $\bm{\gamma} \in \mathbb{R}^{N_2 \times (N_2 - r_2)}$, to the factor matrices $\mathbf{U}_{1}$ and $\mathbf{U}_{2}$, respectively.
These orthogonal complements are the null spaces for the coefficient tensor when matricized over the first and second dimensions respectively.
In other words, they satisfy $\bm{\delta}^\top \mathbf{U}_{1} = \bm{\delta}^\top \mathbf{A}_{(1)} = \bm{0}$ and $\bm{\gamma}^\top \mathbf{U}_{2} = \bm{\gamma}^\top \mathbf{A}_{(2)} = \bm{0}$, allowing us to remove these effects from our model.
Combining the two restrictions in the vectorized model \eqref{eq:rrmar1} gives us $(\bm{\gamma} \otimes \bm{\delta})^\top (\mathbf{U}_2 \otimes \mathbf{U}_1) = (\bm{\gamma} \otimes \bm{\delta})^\top \mathbf{A}_{[2]} = \bm{0}$, amounting to a Kronecker restriction on the null space matrix of the RR-VAR coefficient.

\begin{figure}
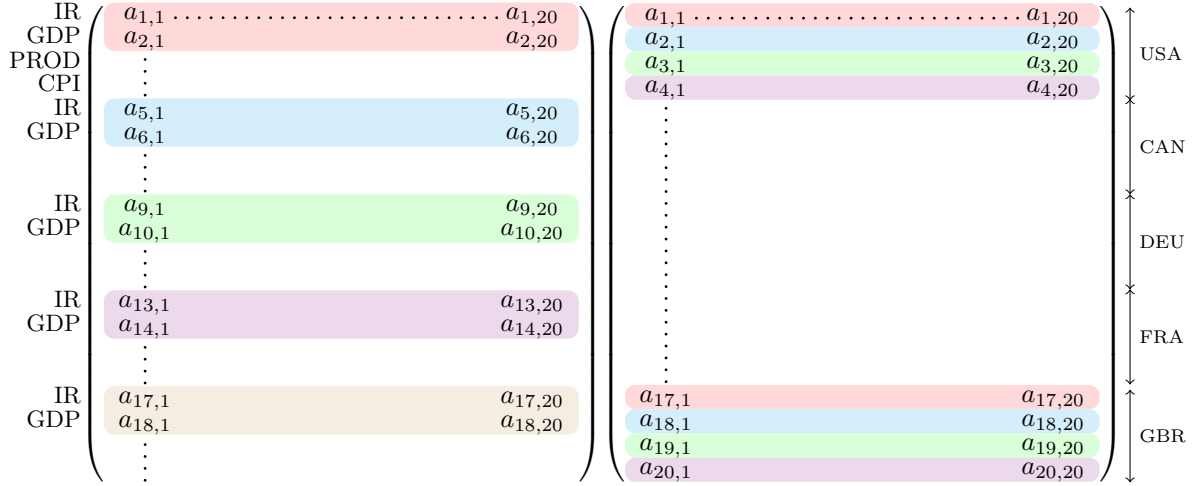

 \[
 \setlength\arraycolsep{5.5pt}
 \renewcommand{\arraystretch}{0.75}
 \begin{pNiceMatrix}%
 [   margin,
     last-col,
     first-col,
   ]
 \text{\small IR} & \Block[fill=red!15,rounded-corners]{2-12}{} a_{1,1} & \Hdotsfor{10} & a_{1,20} & \\
 \text{\small GDP} & a_{2,1} & & & & & & & & & & & a_{2, 20} & \\
 \text{\small PROD} & \Vdotsfor{2} & & & & & & & & & & & & \\
 \text{\small CPI} & & & & & & & & & & & & & \\
 \text{\small IR} & \Block[fill=cyan!15,rounded-corners]{2-12}{} a_{5,1} & & & & & & & & & & & a_{5,20} & \\
 \text{\small GDP} & a_{6,1} & & & & & & & & & & & a_{6,20} & \\
 & \Vdotsfor{2} & & & & & & & & & & & & \\
 & & & & & & & & & & & & & \\
 \text{\small IR} & \Block[fill=green!15,rounded-corners]{2-12}{} a_{9,1} & & & & & & & & & & & a_{9,20} & \\
 \text{\small GDP} & a_{10,1} & & & & & & & & & & & a_{10,20} & \\
 & \Vdotsfor{2}& & & & & & & & & & & & \\
 & & & & & & & & & & & & & \\
 \text{\small IR} & \Block[fill=violet!15,rounded-corners]{2-12}{} a_{13, 1} & & & & & & & & & & & a_{13, 20} & \\
 \text{\small GDP} & a_{14, 1} & & & & & & & & & & & a_{14, 20} & \\
 & \Vdotsfor{2} & & & & & & & & & & & & \\
 & & & & & & & & & & & & & \\
 \text{\small IR} & \Block[fill=brown!15,rounded-corners]{2-12}{} a_{17,1} & & & & & & & & & & & a_{17,20} & \\
 \text{\small GDP} & a_{18,1} & & & & & & & & & & & a_{18,20} & \\
 & \Vdotsfor{2}& & & & & & & & & & & & \\
 & & & & & & & & & & & & & \\
 \end{pNiceMatrix}
 \NiceMatrixOptions{xdots={horizontal-labels,line-style = <->}}
 \begin{pNiceMatrix}%
 [   margin,
     last-col,
   ]
 \Block[fill=red!15,rounded-corners]{1-12}{} a_{1,1} & \Hdotsfor{10}[line-style = standard] & a_{1,20} & \Vdotsfor{4}^{\text{USA}} \\
 \Block[fill=cyan!15,rounded-corners]{1-12}{} a_{2,1} & & & & & & & & & & & a_{2,20}& \\
 \Block[fill=green!15,rounded-corners]{1-12}{} a_{3,1} & & & & & & & & & & & a_{3,20} & \\
 \Block[fill=violet!15,rounded-corners]{1-12}{} a_{4,1} & & & & & & & & & & & a_{4,20} & \\
 \Vdotsfor{12}[line-style = standard] & & & & & & & & & & & & \Vdotsfor{4}^{\text{CAN}} \\
 & & & & & & & & & & & & \\
 & & & & & & & & & & & & \\
 & & & & & & & & & & & & \\
 & & & & & & & & & & & & \Vdotsfor{4}^{\text{DEU}} \\
 & & & & & & & & & & & & \\
 & & & & & & & & & & & & \\
 & & & & & & & & & & & & \\
 & & & & & & & & & & & & \Vdotsfor{4}^{\text{FRA}} \\
 & & & & & & & & & & & & \\
 & & & & & & & & & & & & \\
 & & & & & & & & & & & & \\
 \Block[fill=red!15,rounded-corners]{1-12}{} a_{17,1} & & & & & & & & & & & a_{17,20} & \Vdotsfor{4}^{\text{GBR}} \\
 \Block[fill=cyan!15,rounded-corners]{1-12}{} a_{18,1} & & & & & & & & & & & a_{18,20} & \\
 \Block[fill=green!15,rounded-corners]{1-12}{} a_{19,1} & & & & & & & & & & & a_{19,20} & \\
 \Block[fill=violet!15,rounded-corners]{1-12}{} a_{20,1} & & & & & & & & & & & a_{20,20} & \\
 \end{pNiceMatrix}\]
     \caption{SCCF restriction for $\mathbf{A}_{[2]}$ either in the indicator dimension or the state dimension. The left matrix shows the restrictions along the indicator dimension (rank $r_1=3$), while the right matrix shows the restrictions along the state dimension (rank $r_2=4$).}
     \label{fig:sccfmat}
 \end{figure}

To illustrate this reduced-rank restriction, we reconsider the case of four economic indicators corresponding to five countries.
In line with our empirical illustration in Section \ref{sec:empirical}, we denote the countries as the United States (USA), Canada (CAN), Germany (DEU), France (FRA), and Great Britain (GBR), and the economic indicators as the interest rate (IR), gross domestic product (GDP), manufacturing production (PROD), and consumer price index (CPI).
This setup entails observing a $N_1 \times N_2 = 4 \times 5$ matrix-valued time series over time.
The resulting four-dimensional coefficient tensor, with dimensions of $4 \times 5 \times 4 \times 5$, transforms a two-dimensional $20 \times 20$ VAR coefficient matrix.
This transformation is achieved through a matricization over the first two dimensions, as illustrated in Figure \ref{fig:sccfmat}.

Our focus is first on the left matrix of Figure \ref{fig:sccfmat}.
This matrix illustrates a toy example of reducing the rank of the economic indicator dimension of the coefficient tensor to three, hence $r_1 = 3$, and leaving the country dimension at full rank with $r_2=5$.
The rank of three implies one SCCF co-movement relation for all four economic indicators.
In our toy example, we let the IR and GDP of each country co-move (as highlighted through the same coloring for each of the five countries in Figure \ref{fig:sccfmat}, left panel), but PROD and CPI are unrestricted (i.e.\ unrestricted elements are not displayed in Figure \ref{fig:sccfmat}, left panel).
This is thus, for simplicity, a stylized example where the single co-movement relation is an SCCF co-movement among only IR and GDP. 
We can quantify this relationship with $a_{1, i} = \delta^* a_{2, i}$, $a_{5, i} = \delta^* a_{6, i}$, $a_{9, i} = \delta^* a_{10, i}$, $a_{13, i} = \delta^* a_{14, i}$, and $a_{17, i} = \delta^* a_{18, i}$ for $i = 1, \dots, 20$ and where $\delta^*$ 
is an arbitrary constant representing the scale with respect to the IR.
In this toy example, we thus capture within-country co-movements, where the IR co-moves with the GDP for each country.

On the other hand, we can have a low rank in the country dimension, as shown on the right matrix of Figure \ref{fig:sccfmat}.
This would represent a rank of four in the country dimension, hence $r_2 = 4$ (and leaving the indicator dimension at full rank with $r_1=4$). We consider a toy example where the USA and GBR move together (as highlighted through the same coloring for each of the four indicators in Figure \ref{fig:sccfmat}, right panel), and CAN, DEU, and FRA remaining unrestricted (and are therefore not displayed in Figure \ref{fig:sccfmat}, right panel).
We can again quantify this relationship by saying $a_{1, i} = \gamma^* a_{17, i}$, $a_{2, i} = \gamma^* a_{18, i}$, $a_{3, i} = \gamma^* a_{19, i}$, and $a_{4, i} = \gamma^* a_{20, i}$ for $i = 1, \dots, 20$ and where $\gamma^*$ 
is an arbitrary constant representing the scale with respect to the first country.
Thus, USA and GBR co-move with one another by the scale of $\gamma^*$.
CAN, DEU, and FRA, on the other hand, are not bound to move in tandem with any other country, allowing them to move freely in our toy example.
These dynamics thus represent instances of across-country co-movements.

In practice, it is difficult to identify the contemporaneous co-movements from the coefficient tensor itself, thus we resort to using the orthogonal complement of the matricized coefficient tensor.
This approach allows us to directly identify the extent to which one series co-moves with another, and compare the magnitude of different series compared to a base series.
To illustrate this, reconsider the first example of within-country co-movements with $r_1=3$ as visualized in the left panel of Figure \ref{fig:sccfmat}.
There exists an orthogonal complement vector $\bm{\delta} \in \mathbb{R}^{4 \times 1}$ such that $\bm{\delta}^\top \mathbf{Y}_t = \bm{\delta}^\top \mathbf{E}_t,$ where $\bm{\delta}^\top \mathbf{E}_t$ is white noise.
This yields five relationships corresponding to each of the five countries in our toy example.
By normalizing the first element of $\bm{\delta}$ to one, we obtain
\begin{align} \label{eq:toy-delta-eqs}
    \bm{\delta}^{*\top} \mathbf{Y}_t =
    \begin{bmatrix}
        y_{11t} + \delta_2^* y_{21t} + \delta_3^* y_{31t} + \delta_4^* y_{41t} \\
        y_{12t} + \delta_2^* y_{22t} + \delta_3^* y_{32t} + \delta_4^* y_{42t} \\
        y_{13t} + \delta_2^* y_{23t} + \delta_3^* y_{33t} + \delta_4^* y_{43t} \\
        y_{14t} + \delta_2^* y_{24t} + \delta_3^* y_{34t} + \delta_4^* y_{44t} \\
        y_{15t} + \delta_2^* y_{25t} + \delta_3^* y_{35t} + \delta_4^* y_{45t}
    \end{bmatrix}^\top,
\end{align}
where $\bm{\delta}^{*\top} = (1, \delta_2^*, \delta_3^*, \delta_4^*)$ is the normalized orthogonal complement vector with respect to the first element of $\bm{\delta}$.
This normalization provides the contemporaneous co-movement relations with respect to the first economic indicator (IR).
For instance, the co-movement relation for the economic indicators of the USA given in equation \eqref{eq:toy-delta-eqs} would be $y_{11t} = -\delta_2^* y_{21t} - \delta_3^* y_{31t} - \delta_4^* y_{41t}$ plus a white noise process represented by the first element of $\bm{\delta}^\top \mathbf{E}_t$.
To connect back to our toy example, the scaling term for the USA GDP to recover the USA IR series would be $-\delta_2^*$, giving us the 
co-movement relation between USA IR and USA GDP.
However, it should be noted that this co-movement relation applies to all countries jointly, with $\delta_2^*, \delta_3^*$, and $\delta_3^*$ representing the scaling terms for all countries in the sample.
In our toy example, we can remove the serially correlated component through a linear combination of USA GDP and USA IR, indicative of a serial correlation common feature of the model.

A similar interpretation applies to the second example of between-country co-movement relations.
By post-multiplying $\mathbf{Y}_t$ by $\bm{\gamma}$, we obtain the scaling terms corresponding to the countries in our example.
It is important to note, however, that in practice, the coefficients $\bm{\delta}^*$ and $\bm{\gamma}^*$ would all be estimated as nonzero, meaning that all indicators may generally contribute to the co-movement relations.

\subsubsection{Relation to Factor Models}
\label{sec:factormodels}

The RR-MAR also exhibits a close relation with both static \citep{connor1993test, bai2002determining} and dynamic factor models  \citep{forni2001dynfactor, stock2002forecasting, stock2002indexes}.
The static factor model can be expressed as
\begin{align*}
    \mathbf{y}_t = \bm{\Lambda} \mathbf{f}_t + \varepsilon_t,
\end{align*}
where $\mathbf{y}_t \in \mathbb{R}^{N}$ is a vector valued time series, $\mathbf{f}_t \in \mathbb{R}^r$ are latent factors with $r \ll N$, $\bm{\Lambda} \in \mathbb{R}^{N \times r}$ is the factor loading matrix, and $\varepsilon_t \in \mathbb{R}^N$ is a random noise term.
Normalization restrictions typically require that $\mathbf{F}^\top \mathbf{F} / T = \mathbf{I}_r$ and that $\bm{\Lambda}^\top \bm{\Lambda}$ forms a full rank diagonal matrix, where $\mathbf{F} = (\mathbf{f}_1, \dots, \mathbf{f}_T)^\top$.

Recall the vectorized RR-MAR($1$) has the form
\begin{align*}
    \text{vec}(\mathbf{Y}_t) &= (\mathbf{U}_{2} \otimes \mathbf{U}_{1}) \mathbf{G}_{[2]} (\mathbf{U}_{4} \otimes \mathbf{U}_{3})^\top \text{vec}(\mathbf{Y}_{t-1}) + \text{vec}(\mathbf{E}_t),
\end{align*}
where we can use Kronecker product rules to reduce the model to a lagged matrix-valued factor model
\begin{align}
    \label{eq:rrmarfac}
    \text{vec}(\mathbf{Y}_t) &= (\mathbf{U}_{2} \otimes \mathbf{U}_{1}) \mathbf{G}_{[2]} \text{vec}(\mathbf{F}_{t-1}) + \text{vec}(\mathbf{E}_t),
\end{align}
where $\mathbf{F}_{t-1} = \mathbf{U}_3^\top \mathbf{Y}_{t-1} \mathbf{U}_4$.
This differs slightly from traditional factor models in the literature as it exclusively incorporates lags of the ``factors'' $\mathbf{F}_t$ whereas traditional factor models include contemporaneous effects.
Indeed, it is important to note that traditional factor models typically assume the factors and the idiosyncratic components are uncorrelated at any lead and lag.
In contrast, in our model, $E(\text{vec}(\mathbf{E}_{t+k}) \text{vec}(\mathbf{F}_t)^\top) = \bm{0}$ for $k > 0$.

We may also represent the RR-MAR to include dynamics to the factors, where the system is being driven by a small autoregressive model of factors, similar to those of \cite{cubadda2022dimension, wang2022tensorvar, wang2024high}, with $\mathcal{G}$ holding the coefficients for this system.
This can be seen through a transformation of equation \eqref{eq:rrmarfac}, where we pre-multiply by $(\mathbf{U}_2 \otimes \mathbf{U}_1)^\top$ to obtain ``response factors'' $(\mathbf{U}_2 \otimes \mathbf{U}_1)^\top \text{vec}(\mathbf{Y}_t) = \text{vec}(\mathbf{U}_1^\top \mathbf{Y}_t \mathbf{U}_2) = \text{vec}(\mathbf{F}_t^{resp})$.
Similarly, we label the lagged factors as ``predictor factors'' through $(\mathbf{U}_4 \otimes \mathbf{U}_3)^\top \text{vec}(\mathbf{Y}_{t-p}) = \text{vec}(\mathbf{F}_{t-p}^{pred})$.
Each transformation projects dependent and independent variables onto a lower dimensional factor space.
The resulting autoregressive model takes the form
\begin{align*}
    \mathbf{F}_t^{resp} &= \sum_{j = 1}^p \mathcal{G}_i \bar{\times}_2 \mathbf{F}_{t-j}^{pred} + \mathbf{W}_t, \\
    \text{vec}(\mathbf{F}_t^{resp}) &= \sum_{j = 1}^p \mathbf{G}_{[2], j} \text{vec}(\mathbf{F}_{t-j}^{pred}) + \text{vec}(\mathbf{W}_t),
\end{align*}
where $\mathbf{W}_t = \mathbf{U}_1^\top \mathbf{E}_t \mathbf{U}_2$ embeds a dynamic pattern and $\mathcal{G}_i$ are the coefficients obtained for each lag as a result of the identity matrix in equation \eqref{eq:rrmarp}.
The vectorized model is exactly the lower dimensional VAR of factors described in \cite{cubadda2022dimension} without the restriction that the row and column spaces are equal.

\section{Estimation and Selection of Tucker Ranks}
\label{sec:estimationselection}

This section begins with a description of the estimator and method of estimation for the RR-MAR model (Section \ref{sec:estimation}).
Additionally, we describe how to determine the Tucker ranks in practice (Section \ref{sec:rankselection}).

\subsection{Estimation of RR-MAR for Fixed Ranks}
\label{sec:estimation}
Consider the estimation of the RR-MAR($p$) model in equation \eqref{eq:rrmarp}.
Assuming the Tucker ranks $r_1, r_2, r_3, r_4$ are known, we define the estimator as
\begin{align}
    \label{eq:objfunc}
    \widehat{\mathcal{A}} &= [[\widehat{\mathcal{G}}; \widehat{\mathbf{U}}_1, \widehat{\mathbf{U}}_2, \widehat{\mathbf{U}}_3, \widehat{\mathbf{U}}_4]] = \nonumber \\
    &\underset{\substack{\mathcal{G} \in \mathbb{R}^{r_1 \times r_2 \times r_3 \times r_4},\\ \mathbf{U}_{i} \in \mathbb{R}^{N_j \times r_i}}}{\arg\min} \left\{ \frac{1}{2T}\sum_{t=1}^T \|\text{vec}(\mathbf{Y}_t) - \left(\mathbf{U}_2 \otimes \mathbf{U}_1\right) \mathbf{G}_{[2]} \left( \mathbf{I}_p \otimes \mathbf{U}_4 \otimes \mathbf{U}_3\right)^\top \text{vec}(\mathbf{X}_{t})\|_F^2\right\},
\end{align}
where $j = 1,2$ and $i = 1,2,3,4$.
Note that the objective function is non-convex (as with many reduced-rank problems), which makes it difficult to solve in practice.
However, gradient descent has proven to work well in the non-convex reduced-rank literature (see e.g., \citealp{chi2019nonconvex, chen2019nonconvex, wang2022common} among others), and
we therefore also adopt this algorithmic approach in our work. 
Details regarding the gradient descent algorithm are relegated to Appendix \ref{sec:gdalg}.

\subsection{Tucker Rank Selection}
\label{sec:rankselection}

Consider the RR-MAR($p$) model given in equation \eqref{eq:mar}.
In practice, the Tucker ranks $r_1, r_2, r_3, r_4$ of the lagged coefficients $\mathcal{A}_i$ are unknown, and it is up to the practitioner to determine them.
We use standard information criteria-- Akaike Information Criteria (AIC, \citealp{akaike1974new}) and Bayesian Information Criteria (BIC, \citealp{schwarz1978estimating}) --to jointly obtain an estimate for the ranks and the number of lags $p$ for the RR-MAR. We therefore label this the rank-lag selection criteria in the remainder of the paper.
Both the AIC and the BIC are defined as
\begin{align*}
    \text{AIC}(r_1, r_2, r_3, r_4, p) = \ln(|\widehat{\Sigma}_{\mathbf{E}}|) + \frac{2}{T} \phi(r_1, r_2, r_3, r_4, p), \\
    \text{BIC}(r_1, r_2, r_3, r_4, p) = \ln(|\widehat{\Sigma}_{\mathbf{E}}|) + \frac{\ln(T)}{T} \phi(r_1, r_2, r_3, r_4, p),
\end{align*}
where $\widehat{\Sigma}_{\mathbf{E}}$ is the covariance of the residuals obtained from estimating the RR-MAR and $\phi(r_1, r_2, r_3, r_4, p)$ is the number of parameters from the Tucker decomposition, given the ranks and the number of lags (see equation \eqref{eq:tuckpars} for more details).
Then the selected ranks and lag order can be obtained based on the minimum value derived from either AIC or BIC.

\begin{remark}
    While there is a vast literature dedicated to finding the ranks, or factors, in multivariate systems, we justify our decision of information criteria compared to alternatives.
    First, while there exist classical techniques such as canonical correlation analysis \citep{velu1986reduced, tiao1989model}, to obtain a test statistic based on the chi-squared distribution, these break down as the number of variables increases.
    Thus, such an approach is not suited for our medium-sized setup which involves studying 20-40 series simultaneously, each with multiple numbers of lags.
    Second, and in contrast to the first, the framework of \cite{lam2012factor} directly addresses the problem of dimensionality.
    This method involves extracting a few factors from a large collection of series and works best as the number of series diverges.
    By using a ratio-type estimate, \cite{lam2012factor} find a ``kink'' in the data where the low-rank structures are most likely to occur.
    However, our interest lies in medium-sized settings with partial or full ranks.
    The ratio estimate for the ranks can not detect a full rank, as it will never choose the maximum number of ranks.
    Yet, full ranks are realistic options in our medium-sized settings and should thus be detectable.
    We, therefore, resort to information criteria as an informal specification test to determine whether a rank reduction is warranted, while also incorporating the selection of an optimal lag structure for the RR-MAR model.
    Note finally that we consider stationary time series and the presence of cointegration within, say a reduced rank matrix error correction model is out of the scope of our paper. 
\end{remark}

\section{Simulation Study}
\label{sec:simulation}

We conduct a simulation study to investigate the performance of our rank selection criteria in selecting the true ranks of the coefficient tensor.
We consider two cases of matrix-valued time series: one with dimensions $N_1 \times N_2 = 4 \times 3$ and the other with dimensions $N_1 \times N_2 = 9 \times 2$.
Data are generated from the RR-MAR($1$) model in equation \eqref{eq:mar}.
The entries of the core tensor $\mathcal{G}$ and the factor matrices $\mathbf{U}_1, \mathbf{U}_2, \mathbf{U}_3$, and $\mathbf{U}_4$ are independently sampled from a standard normal distribution and $\text{vec}(\mathbf{E}_t) \overset{iid}{\sim} N(0,\mathbf{I}_{N_1 N_2})$.
The coefficient tensor $\mathcal{A}$ is thus of size $N_1 \times N_2 \times N_1 \times N_2$ and, when matricized, forms a VAR coefficient matrix of size $N_1 N_2 \times N_1 N_2$.
To maintain the latter's stability condition, we scale the Frobenious norm of the core tensor to four. 
Additionally, we orthogonalize the factor matrices by computing their leading singular vectors.

In the first simulation case with dimensions $N_1 = 4$ and $N_2 = 3$, we consider four Tucker rank settings, (i) fully reduced ($r_1 = r_2 = r_3 = r_4 = 1$), (ii) partially reduced first dimension ($r_1 = 1, r_2 = 3, r_3 = 1, r_4 = 3$), (iii) partially reduced second dimension ($r_1 = 4, r_2 = 1, r_3 = 4, r_4 = 1$), and (iv) no rank reduction ($r_1 = 4, r_2 = 3, r_3 = 4, r_4 = 3$).
In the second simulation case  with a matrix-valued time series of dimensions $N_1 = 9$ and $N_2 = 2$, 
we  consider four similar settings for the Tucker ranks; either (i) fully reduced ($r_1 = r_2 = r_3 = r_4 = 1$), (ii) partially reduced first dimension ($r_1 = 1, r_2 = 2, r_3 = 1, r_4 = 2$), (iii) partially reduced second dimension ($r_1 = 9, r_2 = 1, r_3 = 9, r_4 = 1$), and (iv) no rank reduction ($r_1 = 9, r_2 = 2, r_3 = 9, r_4 = 2$).
These settings allow us to investigate the performance of the information criteria under various reduced-rank settings typically underexplored in the reduced-rank or factor model literature, namely, a partially reduced or full rank.
Across all settings, we generate $T+50$ observations from the RR-MAR($1$) model, using the first $50$ as burn-in, taking $T=100$ and $T=500$, and maintaining a fixed signal-to-noise ratio\footnote{Defined as the maximum eigenvalue of the matricized coefficient tensor over the maximum eigenvalue of the error covariance matrix.} of $0.7$.

To evaluate the performance of our rank selection procedures, we calculate the average rank selected by the information criteria (Average Rank) per dimension, its corresponding standard deviation (Std. Rank), and the frequency of correctly chosen ranks (Freq.\ Correct).
Results are summarized in Tables \ref{tab:sim43} and \ref{tab:sim92} for the two simulation cases respectively.

For the case with $N_1 = 4$ and $N_2 = 3$ in Table \ref{tab:sim43}, ranks are correctly selected at a high rate for the BIC, while AIC performs only marginally worse.
With a true rank of $(1,1,1,1)$, AIC selects the correct rank approximately $70$\%
In contrast, BIC selects the correct rank nearly perfectly for $T=100$, increasing to exactly $100$\%
For partially reduced ranks over either the first or second dimension, AIC correctly selects the rank over $90$\%
BIC consistently selects the correct rank, regardless of the number of observations.
Finally, when the true rank is full, both the AIC and BIC select the correct rank across all simulation settings. 
observe some difference compared to Table \ref{tab:sim43}.
For fully reduced ranks, AIC correctly selects $r_1$ and $r_3$ with a frequency of approximately $94$\%
However, when it comes to $r_2$ and $r_4$, the performance of the AIC reduces slightly, with a frequency correct of $75$\%
The reduction may be attributed to the second and fourth dimensions of the coefficient tensor corresponding to the $N_2 = 2$ dimension of the matrix-valued time series, as opposed to the $N_1 = 9$ dimension, which has a more severe rank reduction.
Nevertheless, AIC's performance on the second and fourth ranks improves as the number of observations grows.
BIC exhibits a similar pattern to AIC regarding the correct selection of ranks $r_1, r_3$ compared to ranks $r_2, r_4$.
However, BIC generally outperforms AIC, selecting the true rank in over $94$\%
For the partially reduced-rank settings, the same overall conclusions hold for Table \ref{tab:sim92} as compared to Table \ref{tab:sim43}. 
Notably, when the true ranks are $(9,1,9,1)$, AIC outperforms BIC for the first time, always selecting the correct ranks.
In contrast, BIC requires sufficient observations with $T=500$ to perfectly select the true ranks.
The decline in the performance of BIC compared to AIC is likely due to its overpenalization of model complexity for smaller sample sizes, leading to a preference for lower ranks rather than the correct full ranks.
Nonetheless, across the entire simulation study, BIC demonstrates superior performance over AIC in correctly selecting the true Tucker ranks.

\begin{landscape}
\setlength{\tabcolsep}{1em} 
\renewcommand{\arraystretch}{1.5} 
\begin{table}[ht]
  \centering
  \caption{Simulation Study with matrix-valued time series of dimension $N_1 = 4$ and $N_2  = 3$. Rank selection with AIC or BIC for $T = 100$ and $T = 500$ observations.}
  \label{tab:sim43}
  \begin{tabular}{
    l
    >{\centering\arraybackslash}p{3.4cm}
    >{\centering\arraybackslash}p{3.4cm}
    >{\centering\arraybackslash}p{3.4cm}
    >{\centering\arraybackslash}p{3.4cm}
    >{\centering\arraybackslash}p{3.4cm}
  }
    \toprule
    \textbf{True Rank} & \textbf{Method} &\textbf{Average Rank} & \textbf{Std. Rank} & \textbf{Freq. Correct} \\
    \midrule
    (1,1,1,1) & AIC (100) & (1.43, 1.42, 1.60, 1.54) & (0.81, 0.71, 1.01, 0.78) & (0.72, 0.71, 0.69, 0.64) \\
    & BIC (100) & (1.00, 1.01, 1.00, 1.01) & (0.00, 0.12, 0.00, 0.11) & (1.00, 0.99, 1.00, 0.99) \\
    & AIC (500) & (1.31, 1.35, 1.41, 1.42) & (0.72, 0.67, 0.87, 0.71) & (0.80, 0.76, 0.78, 0.71) \\
    & BIC (500) & (1.00, 1.00, 1.00, 1.00) & (0.00, 0.00, 0.00, 0.00) & (1.00, 1.00, 1.00, 1.00) \\
    \hline
    (1,3,1,3) & AIC (100) & (1.04, 3.00, 1.11, 3.00) & (0.22, 0.00, 0.40, 0.00) & (0.96, 1.00, 0.91, 1.00) \\
    & BIC (100) & (1.00, 3.00, 1.00, 3.00) & (0.00, 0.00, 0.00, 0.00) & (1.00, 1.00, 1.00, 1.00) \\
    & AIC (500) & (1.01, 3.00, 1.04, 3.00) & (0.10, 0.00, 0.23, 0.00) & (0.99, 1.00, 0.96, 1.00) \\
    & BIC (500) & (1.00, 3.00, 1.00, 3.00) & (0.00, 0.00, 0.00, 0.00) & (1.00, 1.00, 1.00, 1.00) \\
    \hline
    (4,1,4,1) & AIC (100) & (4.00, 1.01, 4.00, 1.05) & (0.00, 0.10, 0.00, 0.25) & (1.00, 0.99, 1.00, 0.96) \\
    & BIC (100) & (4.00, 1.00, 4.00, 1.00) & (0.00, 0.00, 0.00, 0.00) & (1.00, 1.00, 1.00, 1.00) \\
    & AIC (500) & (4.00, 1.00, 4.00, 1.02) & (0.00, 0.04, 0.00, 0.15) & (1.00, 0.99, 1.00, 0.98) \\
    & BIC (500) & (4.00, 1.00, 4.00, 1.00) & (0.00, 0.00, 0.00, 0.00) & (1.00, 1.00, 1.00, 1.00) \\
    \hline
    (4,3,4,3) & AIC (100) & (4.00, 3.00, 4.00, 3.00) & (0.00, 0.00, 0.00, 0.00) & (1.00, 1.00, 1.00, 1.00) \\
    & BIC (100) & (4.00, 3.00, 4.00, 3.00) & (0.00, 0.00, 0.00, 0.00) & (1.00, 1.00, 1.00, 1.00) \\
    & AIC (500) & (4.00, 3.00, 4.00, 3.00) & (0.00, 0.00, 0.00, 0.00) & (1.00, 1.00, 1.00, 1.00) \\
    & BIC (500) & (4.00, 3.00, 4.00, 3.00) & (0.00, 0.00, 0.00, 0.00) & (1.00, 1.00, 1.00, 1.00) \\
    \bottomrule
  \end{tabular}
\end{table}
\end{landscape}

\begin{landscape}
\setlength{\tabcolsep}{1em} 
\renewcommand{\arraystretch}{1.5} 
\begin{table}[ht]
  \centering
  \caption{Simulation Study with matrix-valued time series of dimension $N_1 = 9$ and $N_2  = 2$. Rank selection with AIC or BIC for $T = 100$ and $T = 500$ observations.}
  \label{tab:sim92}
  \begin{tabular}{
    l
    >{\centering\arraybackslash}p{3.4cm}
    >{\centering\arraybackslash}p{3.4cm}
    >{\centering\arraybackslash}p{3.4cm}
    >{\centering\arraybackslash}p{3.4cm}
    >{\centering\arraybackslash}p{3.4cm}
  }
    \toprule
    \textbf{True Rank} & \textbf{Method} &\textbf{Average Rank} & \textbf{Std. Rank} & \textbf{Freq. Correct} \\
    \midrule
    (1,1,1,1) & AIC (100) & (1.08, 1.25, 1.16, 1.25) & (0.38, 0.43, 0.85, 0.43) & (0.94, 0.75, 0.94, 0.75) \\  
    & BIC (100) & (1.00, 1.06, 1.00, 1.06) & (0.00, 0.24, 0.00, 0.24) & (1.00, 0.94, 1.00, 0.94) \\
    & AIC (500) & (1.07, 1.18, 1.06, 1.19) & (0.30, 0.39, 0.29, 0.39) & (0.94, 0.82, 0.95, 0.81) \\
    & BIC (500) & (1.00, 1.01, 1.00, 1.01) & (0.00, 0.09, 0.00, 0.09) & (1.00, 0.99, 1.00, 0.99) \\
    \hline
    (1,2,1,2) & AIC (100) & (1.11, 2.00, 1.18, 2.00) & (0.43, 0.00, 0.81, 0.00) & (0.93, 1.00, 0.92, 1.00) \\
    & BIC (100) & (1.00, 2.00, 1.00, 2.00) & (0.00, 0.00, 0.00, 0.00) & (1.00, 1.00, 1.00, 1.00) \\
    & AIC (500) & (1.06, 2.00, 1.08, 2.00) & (0.31, 0.00, 0.36, 0.00) & (0.95, 1.00, 0.94, 1.00) \\
    & BIC (500) & (1.00, 2.00, 1.00, 2.00) & (0.00, 0.00, 0.00, 0.00) & (1.00, 1.00, 1.00, 1.00) \\
    \hline
    (9,1,9,1) & AIC (100) & (8.97, 1.00, 8.97, 1.00) & (0.19, 0.00, 0.19, 0.04) & (0.97, 1.00, 0.98, 1.00) \\
    & BIC (100) & (8.71, 1.00, 8.71, 1.00) & (0.65, 0.00, 0.65, 0.00) & (0.82, 1.00, 0.82, 1.00) \\
    & AIC (500) & (9.00, 1.00, 9.00, 1.00) & (0.00, 0.00, 0.00, 0.00) & (1.00, 1.00, 1.00, 1.00) \\
    & BIC (500) & (8.99, 1.00, 8.99, 1.00) & (0.09, 0.00, 0.09, 0.00) & (0.99, 1.00, 0.99, 1.00) \\
    \hline
    (9,2,9,2) & AIC (100) & (9.00, 2.00, 9.00, 2.00) & (0.00, 0.00, 0.00, 0.00) & (1.00, 1.00, 1.00, 1.00) \\
    & BIC (100) & (8.83, 2.00, 8.58, 2.00) & (0.38, 0.00, 0.65, 0.00) & (0.83, 1.00, 0.67, 1.00) \\
    & AIC (500) & (9.00, 2.00, 9.00, 2.00) & (0.00, 0.00, 0.00, 0.00) & (1.00, 1.00, 1.00, 1.00) \\
    & BIC (500) & (9.00, 2.00, 9.00, 2.00) & (0.00, 0.00, 0.00, 0.00) & (1.00, 1.00, 1.00, 1.00) \\
    \bottomrule
  \end{tabular}
\end{table}
\end{landscape}

\section{Applications}
\label{sec:empirical}
The following two subsections provide applications of the proposed RR-MAR methodology.
First, we discuss an application involving different economic indicators across North American and Eurozone countries (Section \ref{sec:macroinds}).
Next, we examine an application focusing on coincident and leading indicators among U.S.\ states (Section \ref{sec:coincidentleading}). 

\subsection{Macroeconomic indicators for various countries}
\label{sec:macroinds}
\begin{figure}
    \centering
    \includegraphics[scale = 0.29]{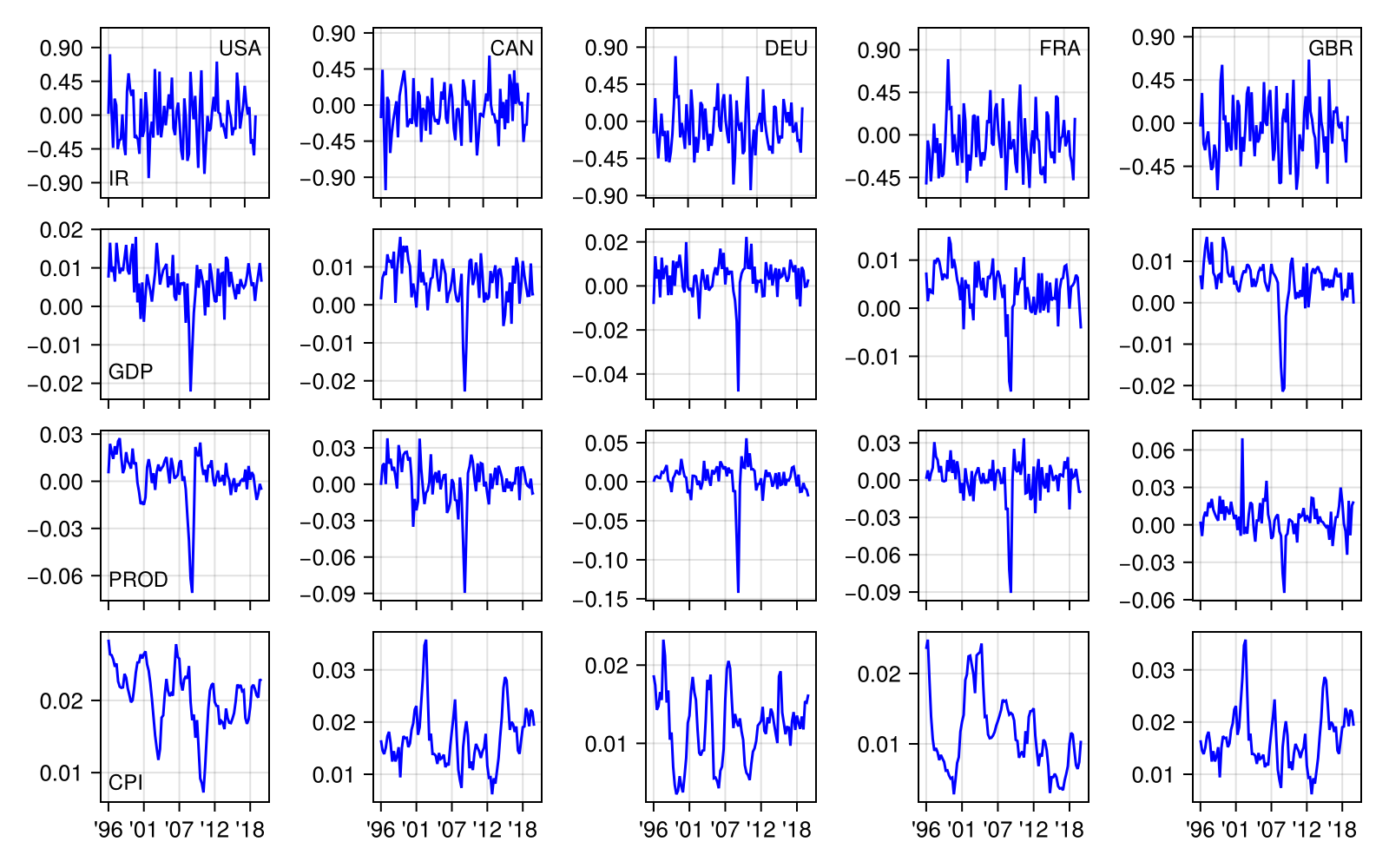}
    \caption{Time series plots for macroeconomic economic indicators (rows) and countries (columns).}
    \label{fig:globaldata}
\end{figure}

We first consider an application inspired by \cite{chen2021mar}, who examine data from $N_1 = 4$ macroeconomic indicators over $N_2 = 5$ countries.
The dataset consists of quarterly observations spanning the first quarter of 1996 to the fourth quarter of 2019, totaling $T = 96$ observations.
The four economic indicators are long-term interest rates (IR) (government bonds maturing in ten years), real Gross Domestic Product (GDP), manufacturing production (PROD), and inflation measured by the Consumer Price Index (CPI).
The five countries include the United States (USA), Canada (CAN), Germany (DEU), France (FRA), and the United Kingdom (GBR).
The data is obtained from the Organization for Economic Co-operation and Development (OECD) at \url{https://data-explorer.oecd.org/}.
For data transformations, we take the first differences of the interest rates, while GDP and manufacturing production are the logged differences.
For CPI we take the annual inflation series.
The transformed series for the four indicators and five countries are visualized in Figure \ref{fig:globaldata}.

Our aim is to investigate the co-movement structures among the indicators and/or countries. To this end, we compute our rank-lag selection criteria, AIC and BIC, with a maximum lag of three.
The results point toward a reduced-rank structure along both economic indicator and country dimensions: AIC selects Tucker ranks $(4,5,4,5)$ with three lags, hence no rank reduction along either dimension.  BIC, on the other hand, selects Tucker ranks $(3,1,4,3)$ with one lag, thereby signaling a reduced-rank structure along the indicator and country dimensions.
Given the excellent performance of the BIC in our simulation study, we continue our empirical analysis with the Tucker ranks selected by the BIC.

{\bf Serial Correlation Common Features.} 
We first consider the SCCF restrictions, the Tucker ranks $r_1 = 3$ and $r_2 = 1$ can then be used to discuss the implications for the economic indicator and country dimension in more detail.
The ranks are indicative of a partially reduced economic indicator dimension, while the country dimension is fully reduced.
Next, we take the null space for the factor matrices $\mathbf{U}_1$ and $\mathbf{U}_2$ and subsequently normalize it with regard to a single economic indicator or country respectively.
Starting with the economic indicators, $r_1 = 3$ implies a single co-movement relation between all four economic indicators.
Normalizing over the second variable (GDP), the null space vector is $\bm{\delta}^{*\top} = (0.0001, 1, -0.3331, 0.1194)$.
This indicates that GDP movements are approximately one-third of those in manufacturing production across all countries, while the core CPI inversely co-moves with GDP.

For the country dimension, we have a rank of one ($r_2 = 1$), thus all countries co-move.
Suppose we are interested in the co-movements of the U.S., the firstly ordered country. 
Then our normalization entails the null space matrix
\begin{align*}
    \bm{\gamma}^* = \begin{bmatrix}
    -0.85 & -0.74 & -0.79 & -0.77 \\
    1.00 & 0.00 & 0.00 & 0.00 \\
    0.00 & 1.00 & 0.00 & 0.00 \\
    0.00 & 0.00 & 1.00 & 0.00 \\
    0.00 & 0.00 & 0.00 & 1.00 \\
    \end{bmatrix}.
\end{align*}
These results thus indicate clear, almost one-to-one co-movements of all countries with the U.S., with the Canadian economic indicators most closely moving together with the U.S. ones, as evident from the first column.

\begin{figure}
    \centering
    \includegraphics[scale=0.21]{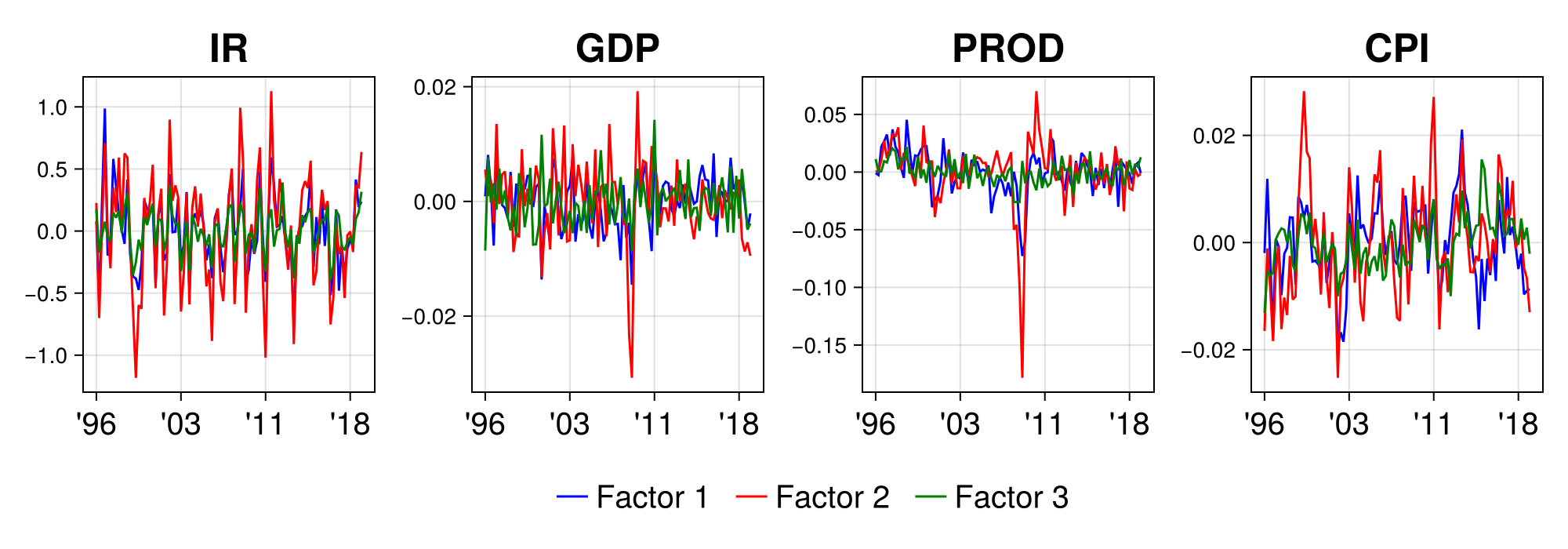}
    \caption{Predictor factors obtained from $\mathbf{U}_3^\top \mathbf{Y}_{t-1} \mathbf{U}_4$ with a full rank in the economic indicator dimension ($r_3=4$, as visualized in the different panels), while the country dimension is reduced to $r_4=3$ country factors (different lines per panel).}
    \label{fig:idfactors}
\end{figure}

{\bf Factor Models.}
Next, we discuss the implications of our framework for factor models 
thereby further inspecting the Tucker ranks $r_3=4$ and $r_4=3$.
We  construct predictor factors using $\mathbf{U}_3$ and $\mathbf{U}_4$, namely as $\mathbf{F}_t^{pred} = \mathbf{U}_3^\top \mathbf{Y}_{t-1} \mathbf{U}_4$.
This results in an $r_3 \times r_4 = 4 \times 3$ matrix-valued time series of factors, analogous to ``indexes'' in the index model literature.
Here, the Tucker ranks $r_3 = 4$ and $r_4 = 3$ define the factor dimensions, representing a full rank economic indicator dimension and a reduced rank country dimension, respectively.
The fact that the economic indicators are of full rank yields an interesting characteristic of the factors, as we know exactly which factor corresponds to each economic indicator.
In contrast, the reduced country dimension ($r_4=3$) implies that the country factors summarize country-specific information for each economic indicator.
Leveraging this, we can obtain country factors for each economic indicator, with the resulting series illustrated in Figure \ref{fig:idfactors}.

\begin{figure}
    \centering
    \includegraphics[scale=0.18]{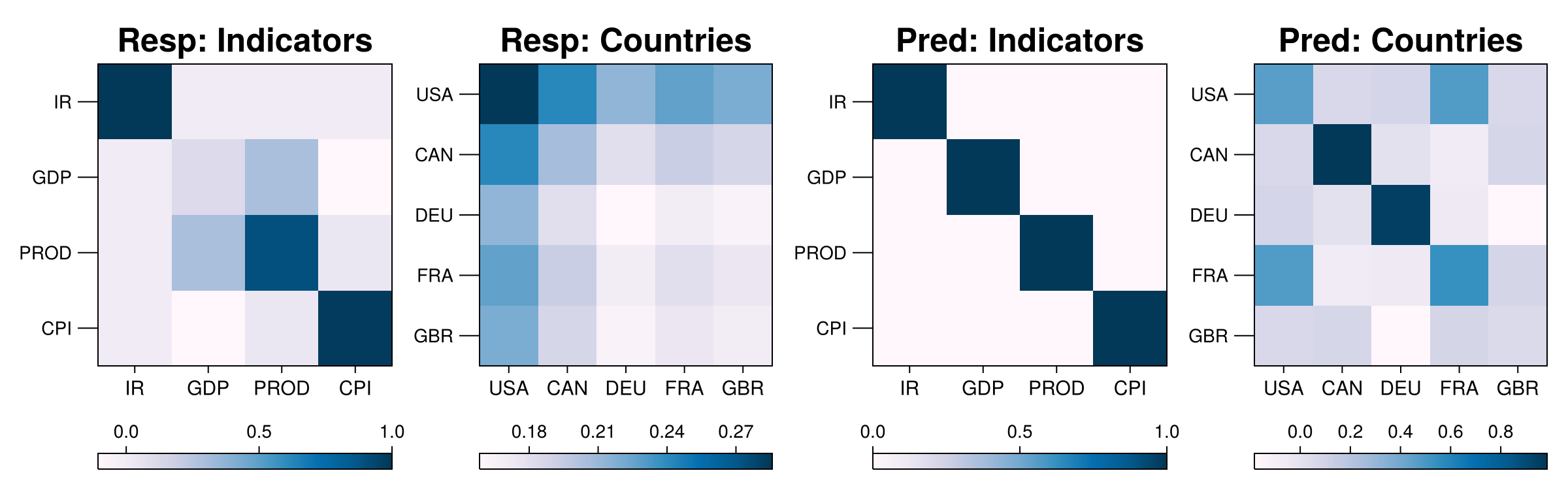}
    \caption{Projection matrices for $\mathbf{U}_1$, $\mathbf{U}_2$, $\mathbf{U}_3$, $\mathbf{U}_4$. Each corresponds to either the response factors or predictor factors for the countries and economic indicators respectively.}
    \label{fig:countryheat}
\end{figure}

Finally, we construct projection matrices to visualize the relationships between economic indicators and countries using predictor and response factors as in \cite{wang2024high}.
These projection matrices are found by computing $\mathbf{U}_i \mathbf{U}_i^\top$ for each dimension of the coefficient tensor $i = 1,2,3,4$, and are exactly identified.
Four projection matrices are presented in Figure \ref{fig:countryheat}, depicting heatmaps of response and predictor factors for countries and economic indicators respectively.
The response factors given in the two left matrices align closely with our SCCF interpretations.
Specifically, from the first panel in Figure \ref{fig:countryheat}, there are seemingly three groupings for the economic indicators: one for interest rates, another for GDP and PROD, and a final group consisting of the CPI.
This can be observed from the darker shades on the diagonal for IR and CPI, while GDP and PROD combined have darker shades.
The second panel in Figure \ref{fig:countryheat} then highlights that 
a single country factor predominantly accounts for the variance of the factors, as given by the darker shade of the USA row and column.

Shifting our focus to the predictor factors given in the right two matrices of Figure \ref{fig:countryheat};  
the identity matrix in the third panel of Figure \ref{fig:countryheat} clearly reveals that the indicators are left unrestricted (full Tucker rank $r_3)$.
The last panel in Figure \ref{fig:countryheat} highlights that the predictor country factors are explained best by Canada and Germany, as given by the darker shades on the diagonal of the heatmap.
USA and FRA show a lighter color but still contribute to the prediction of the countries and indicators in the RR-MAR.
In fact, there seems to be an interplay between the USA and FRA countries, as given by the same shade of blue on the off-diagonals.
Finally, GBR does not enter into the predictor factors for the countries, with a value close to zero on the diagonal of the projection matrix heatmap.

\subsection{Coincident and Leading Indexes among U.S.\ States}
\label{sec:coincidentleading}

\begin{figure}
    \centering
    \includegraphics[scale=0.19]{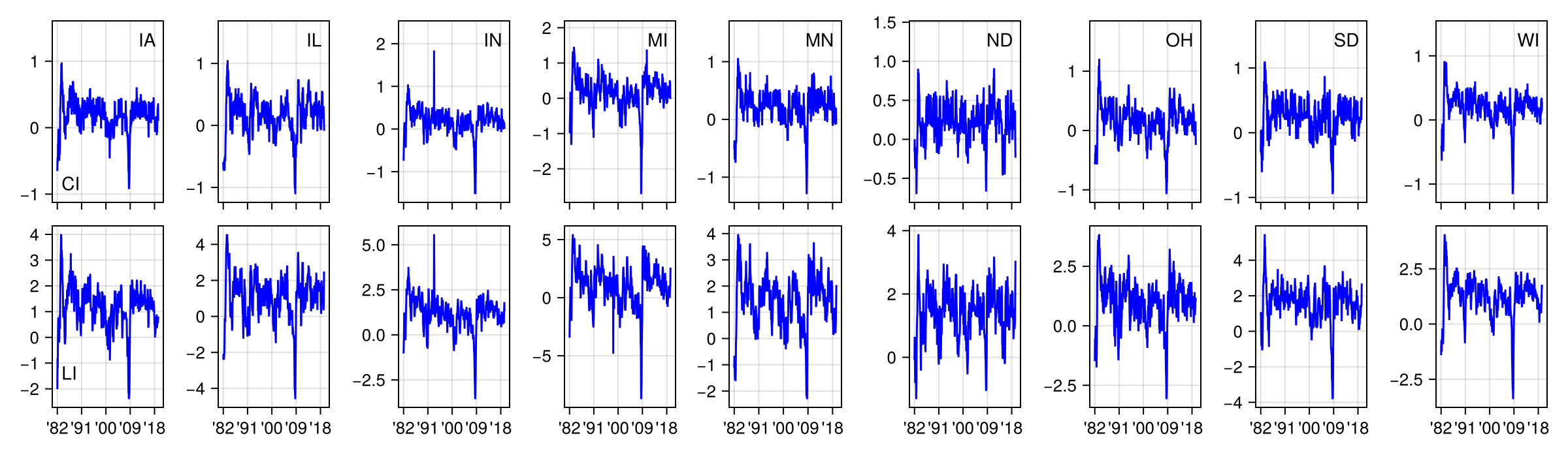}
    \caption{Time series plots of the $2 \times 9$ matrix-valued time series for growth rates of coincident and leading indexes among different North Central U.S. states.}
    \label{fig:stateplots}
\end{figure}

We consider a system with $N_1 = 2$ economic indicators, namely
monthly coincident and leading indexes (referred to as CI and LI respectively),  measured for $N_2 = 9$ North Central U.S. states, namely Illinois (IL), Indiana (IN), Iowa (IA), Michigan (MI), Minnesota (MN), North Dakota (ND), South Dakota (SD), Ohio (OH), and Wisconsin (WI).
The data for these series are sourced from the Federal Reserve Bank of Philadelphia at \url{https://www.philadelphiafed.org/surveys-and-data/regional-economic-analysis/}.

The coincident indexes combine four state-level indicators to summarize prevailing economic conditions.
This includes indicators such as non-farm payroll employment, average hours worked in manufacturing by production workers, the unemployment rate, and wage and salary deflated by the consumer price index.
These four variables are combined through a dynamic single-factor model, as outlined by \cite{stock1989new} into a single coincident index.
The coincident index is seasonally adjusted on the sample spanning from January 1982 to February 2020, excluding the pandemic period, resulting in a sample size of $T=458$ monthly observations.
Furthermore, we compute monthly growth rates for each coincident indicator as visualized in the top panel of Figure \ref{fig:stateplots}.\footnote{For Michigan, we adjust the level value of July 1998 by averaging the values of June and August. This adjustment is made to mitigate the impact of a drop in levels, which otherwise generates two outliers in the first differences of the data.}

The leading indexes, on the other hand, are inherently presented as monthly growth rates.
Each state's leading index anticipates the six-month growth rate of its coincident index.
In addition, the indicators that form the leading index include the coincident index, state-level housing permits, state initial unemployment claims, delivery times from the Institute for Supply Management manufacturing survey, and the interest rate spread between the 10-year Treasury bond and the three-month Treasury bill.
These leading indexes are plotted in the bottom panel of Figure \ref{fig:stateplots}.

For this matrix-valued time series of dimensions $N_1=2$ and $N_2=9$, 
we do not necessarily expect the coincident and leading indicators to co-move contemporaneously.
Indeed, each state's leading index is designed to forecast the six-month growth rate of its coincident index.
The nature of the two indexes is thus different; coincident indicators reflect current economic conditions while leading indicators aim to predict future trends in the coincident index.
Similarly, co-movement between the states may or may not occur since 
neighboring economies are often interconnected and interact with one another.

To investigate the possible presence of co-movement within and between the dimensions of this $2\times 9$ matrix-valued time series, we compute our rank-lag selection criteria with a maximum lag of three.
As a result, AIC selects a full-rank model with two lags, while BIC selects a full-rank model with one lag.
These results thus suggest no evidence for co-movement among the indicators or the countries and highlight the capability of our selection procedure to handle full-rank coefficient tensors when supported by the data.

\section{Conclusion}
\label{sec:conclusion}
This paper proposes a reduced-rank regression for medium-sized matrix-valued time series, denoted as the RR-MAR.
By assuming the coefficient of the regression to be a tensor, we apply a Tucker decomposition on the coefficients to individually reduce the ranks of the different dimensions of the coefficient tensor.
We then relate the rank reductions of the coefficient tensor to two popular reduced-rank methodologies in the literature: the serial correlation common feature and the factor index model.
In doing so, we can detect different co-movement relationships from the rows and columns of the matrix-valued time series.
Furthermore, our framework allows for different ranks in the dimensions of the autoregressive coefficient.
This allows us to investigate the potential of a partial full rank coefficient, where one dimension is of reduced rank and another is unrestricted.
We demonstrate the versatility of our method across various simulation settings and two empirical data sets; one involving economic indicators and world countries where we demonstrate the capability of the proposed rank-lag selection criterion to uncover reduced Tucker ranks, and another with coincident and leading indicators for North Central U.S. states where we demonstrate its ability to uncover full Tucker ranks.
All code and replication material for this paper can be found in the Julia repository \texttt{RR-MAR} on the second author's GitHub page \url{https://github.com/ivanuricardo/RR-MAR}.

The proposed methodology can be extended in several directions.
It would be interesting to consider a further disaggregation of dimensions and create a tensor autoregression instead of a matrix autoregression.
This disaggregation would allow us to analyze more dimensions of a data tensor and to have even more diverse  SCCF or factor model relations between the different dimensions.
Furthermore, while we allow for the inclusion of several lags in the RR-MAR, we do not explore the possibility of a rank reduction in the lag dimension \citep{ahn1988nested}.
\cite{wang2022tensorvar} does this in their VAR setup, but its impact on the factor matrices in an RR-MAR is yet to be explored.
Another interesting avenue for future research concerns nonstationary matrix-valued time series, thereby leading to the exploration of a matrix-valued error correction model.
In this way, a trend common feature could be explored instead of the serial correlation common feature. 

\bigskip
\noindent
{\bf Acknowledgements.}
The last author was financially supported by the Dutch Research Council (NWO) under grant number VI.Vidi.211.032.

\bibliographystyle{asa}
\bibliography{references}
\newpage

\appendix

\section{Appendix}
\label{sec:gdalg}

\begin{algorithm}
\caption{Gradient descent algorithm for RR-MAR}
\label{alg:gd}
\begin{algorithmic}[0]
\State \textbf{Input}: \\
Tucker ranks $(r_1, r_2, r_3, r_4)$, lag order $p$, initialization $\mathcal{G}^{(0)}$ and $\mathbf{U}_i^{(0)}$ for $i = 1, \dots, 4$, step size $\eta = 1e-03$, convergence tolerance $\epsilon < 1e-03$, maximum iterations = $500$.
\Repeat \quad s = 1, 2, \dots
    \State $\mathbf{U}_1^{(s)} \gets \mathbf{U}_1^{(s-1)} - \eta \nabla \mathcal{L}_1(\mathbf{U}_1^{(s-1)}, \mathbf{U}_2^{(s-1)}, \mathbf{U}_3^{(s-1)}, \mathbf{U}_4^{(s-1)}, \mathcal{G}^{(s-1)})$
    \State $\mathbf{U}_2^{(s)} \gets \mathbf{U}_2^{(s-1)} - \eta \nabla \mathcal{L}_2(\mathbf{U}_1^{(s)}, \mathbf{U}_2^{(s-1)}, \mathbf{U}_3^{(s-1)}, \mathbf{U}_4^{(s-1)}, \mathcal{G}^{(s-1)})$
    \State $\mathbf{U}_3^{(s)} \gets \mathbf{U}_3^{(s-1)} - \eta \nabla \mathcal{L}_3(\mathbf{U}_1^{(s)}, \mathbf{U}_2^{(s)}, \mathbf{U}_3^{(s-1)}, \mathbf{U}_4^{(s-1)}, \mathcal{G}^{(s-1)})$
    \State $\mathbf{U}_4^{(s)} \gets \mathbf{U}_4^{(s-1)} - \eta \nabla \mathcal{L}_4(\mathbf{U}_1^{(s)}, \mathbf{U}_2^{(s)}, \mathbf{U}_3^{(s)}, \mathbf{U}_4^{(s-1)}, \mathcal{G}^{(s-1)})$
    \State $\mathcal{G}^{(s)} \gets \mathcal{G}^{(s-1)} - \eta \nabla \mathcal{L}_{\mathcal{G}}(\mathbf{U}_1^{(s)}, \mathbf{U}_2^{(s)}, \mathbf{U}_3^{(s)}, \mathbf{U}_4^{(s)}, \mathcal{G}^{(s-1)})$
\Until{convergence} 
\State \textbf{Finalize}: \\
$\widehat{\mathbf{U}}_i \gets$ top $r_i$ left singular vectors of $\widehat{\mathbf{A}}_{(i)} = [[\mathcal{G}^{(s)}; \mathbf{U}_1^{(s)}, \mathbf{U}_2^{(s)}, \mathbf{U}_3^{(s)}, \mathbf{U}_4^{(s)}, \mathbf{I}_p]]_{(i)}$ \\
$\widehat{\mathcal{G}} \gets \mathcal{A} \times_1 \mathbf{U}_1^\top \times_2 \mathbf{U}_2^\top \times_3 \mathbf{U}_3^\top \times_4 \mathbf{U}_4^\top \times_5 \mathbf{I}_p$
\end{algorithmic}
\end{algorithm}
Algorithm \ref{alg:gd} presents the gradient descent algorithm to solve the optimization problem in equation \eqref{eq:objfunc}.
We begin with an initial estimation of the $\mathcal{G}^{(0)}$ core tensor and the $\mathbf{U}_i^{(0)}$ factor matrices, for $i = 1, \dots 4$.
This can be obtained by performing the HOSVD on the OLS solution of the coefficient $\mathcal{A}$ in equation \eqref{eq:mar}.
Note that many different randomized initialization schemes may be tested for the convergence and performance of the estimate.
We found the OLS initialization to be the best initialization across our numerical experiments.

Once starting values are obtained, the gradient descent iterates may be found via Algorithm \ref{alg:gd}.
For each factor matrix and core tensor, $\nabla \mathcal{L}_i(\mathbf{U}_1, \mathbf{U}_2, \mathbf{U}_3, \mathbf{U}_4, \mathcal{G})$ and $\nabla \mathcal{L}_\mathcal{G}(\mathbf{U}_1, \mathbf{U}_2, \mathbf{U}_3, \mathbf{U}_4, \mathcal{G})$ are functions yielding the tensor representations of the gradient conditions for the OLS solution of the objective function in equation \eqref{eq:objfunc}.
This representation allows for a closed-form solution of the factor matrices and core tensor given by
\begin{align*}
    \label{eq:grad}
    \nabla \mathcal{L}_1(\mathbf{U}_1, \mathbf{U}_2, \mathbf{U}_3, \mathbf{U}_4, \mathcal{G}) &= \mathcal{M}_{(1)} \left( \mathbf{I}_p \otimes \mathbf{U}_4 \otimes \mathbf{U}_3 \otimes \mathbf{U}_2 \right) \mathcal{G}_{(1)}^\top, \\
    \nabla \mathcal{L}_2(\mathbf{U}_1, \mathbf{U}_2, \mathbf{U}_3, \mathbf{U}_4, \mathcal{G}) &= \mathcal{M}_{(2)} \left( \mathbf{I}_p \otimes \mathbf{U}_4 \otimes \mathbf{U}_3 \otimes \mathbf{U}_1 \right) \mathcal{G}_{(2)}^\top, \\
    \nabla \mathcal{L}_3(\mathbf{U}_1, \mathbf{U}_2, \mathbf{U}_3, \mathbf{U}_4, \mathcal{G}) &= \mathcal{M}_{(3)} \left( \mathbf{I}_p \otimes \mathbf{U}_4 \otimes \mathbf{U}_2 \otimes \mathbf{U}_1 \right) \mathcal{G}_{(3)}^\top, \\
    \nabla \mathcal{L}_4(\mathbf{U}_1, \mathbf{U}_2, \mathbf{U}_3, \mathbf{U}_4, \mathcal{G}) &= \mathcal{M}_{(4)} \left( \mathbf{I}_p \otimes \mathbf{U}_3 \otimes \mathbf{U}_2 \otimes \mathbf{U}_1 \right) \mathcal{G}_{(4)}^\top, \\
    \nabla \mathcal{L}_{\mathcal{G}}(\mathbf{U}_1, \mathbf{U}_2, \mathbf{U}_3, \mathbf{U}_4, \mathcal{G}) &= \mathcal{M} \times_{1} \mathbf{U}_1^{\top (s)} \times_2 \mathbf{U}_2^{\top (s)} \times_{3} \mathbf{U}_3^{\top (s)} \times_{4} \mathbf{U}_4^{\top (s)} \times_5 \mathbf{I}_p,
\end{align*}
where
    $\mathcal{M} = T^{-1} \sum_{t=1}^T \left(\mathcal{G} \times_1 \mathbf{U}_1 \times_2 \mathbf{U}_2 \times_3 \mathbf{U}_3 \times_4 \mathbf{U}_4 \times_5 \mathbf{I}_p \right) \bar{\times}_3 \mathcal{X}_{t}-\mathbf{Y}_t \circ \mathcal{X}_{t}$
depends on all factor matrices and core tensor $\mathcal{G}$ while utilizing the stacked lag matrix $\mathcal{X}_t$ as defined in equation \eqref{eq:stackedrrmarp} and $\circ$ denotes the tensor outer product.
The resulting $\mathcal{M}$ is a $N_1 \times N_2 \times N_1 \times N_2 \times p$ tensor, while $\mathcal{M}_{(i)}$ is the matricized representation of this tensor over the $i$th dimension.

\end{document}